\newcommand{\version}{April 22, 2021}
\documentclass[letterpaper,oneside,11pt,pdftex]{article}
\pdfoutput=1
\usepackage[bindingoffset=0.0cm,textheight=22.60cm,hdivide={*,16.3cm,*}, vdivide={*,22.60cm,*}]{geometry}
\usepackage{amsmath,amsfonts,amssymb}
\usepackage{mathtools}
\usepackage[pdftex]{graphicx}
\usepackage[margin=1cm,font=small]{caption}
\usepackage[pdftex]{grffile}
\usepackage[T1]{fontenc}
\usepackage[utf8]{inputenc}
\usepackage{fouriernc}

\usepackage[numbers,square,comma,sort&compress]{natbib}
\usepackage[pdftex,hyperref,svgnames]{xcolor}
\usepackage[pdftex,bookmarksnumbered=true,breaklinks=true,%
colorlinks=true,linktocpage=true,linkcolor=MediumBlue,citecolor=ForestGreen,urlcolor=DarkRed]{hyperref}

\makeatletter
\AtBeginDocument{
  \hypersetup{
    pdfkeywords = {\keywords},
    pdftitle = {\@title},
    pdfauthor = {\@author}
  }
}
\makeatother

\makeatletter

 \@addtoreset{equation}{section}
 \makeatother

\renewcommand{\a}{\alpha}
\renewcommand{\b}{\beta}
\newcommand{\g}{\gamma}

\newcommand{\vth}{\vartheta}

\newcommand{\m}{\mu}

\newcommand{\tr}{\textrm{tr}}
\newcommand{\sgn}[1]{\textrm{sgn}\!\left(#1\right)}
\newcommand{\nn}{\nonumber}
\newcommand{\eqnref}[1]{Eq. \eqref{#1}}

\newcommand{\kb}{k_{\txt{B}}}

\newcommand{\ct}{c_{\textrm{T}}}
\newcommand{\cl}{c_{\textrm{L}}}
\newcommand{\cs}{v_{\textrm{c}}}

\newcommand*{\mat}[1]{\mathbf{#1}}

\newcommand{\txt}[1]{\textrm{#1}}

\DeclarePairedDelimiter\abs{\lvert}{\rvert}
\newcommand{\coleq}{\vcentcolon=}
\newcommand{\figtitle}[1]{\textbf{\emph{\footnotesize{#1}}}}

\title{\texorpdfstring{\begin{flushright}
        {\small LA-UR-20-25546}
       \end{flushright}\vspace{2em}}{}%
       Dislocation drag and its influence on elastic precursor decay}

\author{Daniel N. Blaschke and Darby J. Luscher}

\date{\version}

\newcommand{\keywords}{dislocations in crystals, drag coefficient,  elastic precursor decay}

\graphicspath{{./}{./figures/}}

\begin{document}

\maketitle

\thispagestyle{empty}
\begin{center}
\vspace{-0.3cm}
Los Alamos National Laboratory, Los Alamos, NM, 87545, USA
\\[0.5cm]
\ttfamily{E-mail: dblaschke@lanl.gov, djl@lanl.gov}
\end{center}

\begin{abstract}
Plastic deformation is mediated by the creation and movement of dislocations, and at high stress the latter is dominated by dislocation drag from phonon wind.
By simulating a 1-D shock impact problem we analyze the importance of accurately modeling dislocation drag and dislocation density evolution in the high stress regime.
Dislocation drag is modeled according to a first-principles derivation as a function of stress and dislocation character, and its temperature and density dependence are approximated to the extent currently known.
Much less is known about dislocation density evolution, leading to far greater uncertainty in these model parameters.
In studying anisotropic fcc metals with character dependent dislocations, the present work generalizes similar earlier studies by other authors.
% who considered only edge dislocations in the isotropic approximation.
\end{abstract}

\vspace{1cm}
% \newpage
\tableofcontents
\newpage

\section{Introduction and background}
\label{sec:intro}
%%%%%%%%%%%%%%%%%%%%%%%%%%%%%%%%%%%%%%%%%%%%%%%%%%%%%%%%%

Material strength models applicable in the drag dominated high stress regime have historically been purely phenomenological \cite{Steinberg:1980,Steinberg:1989,Follansbee:1988,PTW:2003} and as such do not perform well outside of the regime they have been calibrated to \cite{Austin:2018}.
However, in order to improve accuracy and especially predictive capability at the mesoscale, it is necessary to include effects of microscopic physics to some degree.
More recently, many approaches have taken this route, aiming at providing a more accurate description of the high stress regime based on microscopic physics of dislocation mobility, see e.g. \cite{Kuksin:2008,Krasnikov:2010,Barton:2011,Hansen:2013,Lloyd:2014JMPS,Hunter:2015,Borodin:2015,Luscher:2016,Austin:2018} and others.
The main obstacle in these recent endeavors has been the uncertainty as to how the dislocation drag coefficient $B$ behaves at high velocities, temperatures and material densities (and thus pressures).
Lacking a first principles derivation, previous authors have assumed a variety of functional forms for $B$ which range from simple choices like $B=\textrm{const.}$
(see e.g. \cite{Kuksin:2008,Hansen:2013,Hunter:2015,Borodin:2015}) to $B\sim\sqrt{v}$ above some threshold velocity \cite{Olmsted:2005,Marian:2006,Cho:2017}, to ``relativistic factors'' $B\sim1/(1-v^2/v_\txt{crit}^2)^m$ with a limiting (critical) velocity $v_\txt{crit}$ and a range of powers $1/2\le m\le 4$ \cite{Krasnikov:2010,Barton:2011,Luscher:2016,Austin:2018};
see also \cite{Gurrutxaga:2016}.
The latter assumptions have their roots in the observation that the dislocation self-energy for a perfect dislocation in the isotropic limit diverges at the transverse sound speed together with dislocation mobility data from molecular dynamics (MD) simulations that indicate an increase in dislocation drag $B$ in the vicinity of the lowest shear wave speed in anisotropic crystals.
Nonetheless, all velocity dependencies mentioned above are proposed ad hoc and lack a first principles motivation:
For example, none of the phenomenological considerations determines the power $m$.
Temperature dependence is typically assumed to be linear (an assumption which has its roots in the linear temperature dependence of the Debye phonon spectrum in its high temperature expansion~\cite{Alshits:1992,Blaschke:BpaperRpt}) with polynomial corrections in some cases \cite{Barton:2011}.
The material density dependence is taken into account only indirectly via the limiting velocity within the relativistic factor, or neglected altogether.
See \cite{Gurrutxaga:2020} for a recent insightful review of high speed dislocation dynamics; for a review of shock compression of crystalline solids and elastic precursor decay, see \cite{Clayton:2018,Clayton:2019book} and references therein.

Recently, a first principles derivation of drag coefficient $B$ based on Refs. \cite{Blaschke:BpaperRpt,Blaschke:2018anis,Blaschke:2019fits,Blaschke:2019Bpap} was incorporated into a strength model in the isotropic limit \cite{Hunter:2015,Blaschke:2019a}, showcasing how $B$ influences stress-strain curves at constant dislocation densities.
Here, we focus on the velocity (or stress) dependence of $B$ and study its influence on a 1-D shock impact problem.
Our method is a single-crystal plasticity simulation along the lines of Refs. \cite{Luscher:2016,Mayeur:2016,Luscher:2017}, but including the first-principles derivation of $B$, or rather  an approximation thereof amenable to in-line computation within our larger simulation.
Furthermore, in contrast to \cite{Luscher:2016,Mayeur:2016} where only edge dislocations were considered, we track dislocations of different characters $\vth$ in this work.
For this purpose we discretize $\vth$ and choose a small number of dislocation character bins.
While most of the calculations adopt two bins of dislocation character, i.e. edge and screw dislocations, we assess the effects of incorporating a finer dislocation character resolution by including up to seven bins in Section \ref{sec:results}.

A main driver for our studies of dislocation drag are speculations in the literature about the cause of the increase in elastic precursor amplitude at high temperatures observed in single crystal and polycrystal aluminum experiments \cite{Kanel:2001,Krasnikov:2011,Zaretsky:2012,Gurrutxaga:2017,Zuanetti:2021}.
Since these experiments pertain to the drag dominated regime, the only temperature dependent quantities that could cause this effect are the dislocation drag and the evolution of mobile dislocation densities within Orowan's relation \cite{Kanel:2001}.
Although it has been speculated earlier in \cite{Zaretsky:2012} that the temperature dependence of the mobile dislocation density may be relevant, the recent work on polycrystal aluminum of Ref. \cite{Austin:2018}, assumed all dislocation density evolution parameters to be temperature independent and tried to explain the increase in elastic precursor amplitude at elevated temperatures solely by a temperature dependent drag coefficient.
The latter was modeled according to MD data for edge dislocations.
In trying to match experimental data on the elastic precursor decay in polycrystalline aluminum from room temperature to melting, discrepancies were found and attributed to temporal resolution limits of the VISAR experiments.

Here, we argue that dislocation drag alone cannot explain this effect, and that an accurate model for temperature dependent dislocation density evolution is needed.
We start by reviewing the single crystal plasticity model employed in this work which is based on Refs. \cite{Luscher:2016,Mayeur:2016,Luscher:2017} in Sections \ref{sec:modelsummary}, \ref{sec:dragcoeff}, and \ref{sec:eos}.
Apart from including a best estimate for the stress, temperature, density, and character dependence of dislocation drag based on Refs. \cite{Blaschke:2018anis,Blaschke:2019Bpap,Blaschke:2019a}, we also improve the dislocation evolution model of \cite{Luscher:2016,Mayeur:2016,Luscher:2017}.
As far as the latter is concerned, however, a good understanding of the temperature dependence of the dislocation evolution behavior is still lacking, but further improvements in this regard are beyond the scope of this paper and are thus left for future work.
The main improvements of the present model are the stress, character, temperature, and density dependent drag coefficient $B$ in \eqnref{eq:BTrho}, as well as the character dependent description of forest dislocations in \eqref{eq:newsource_nonucl}, \eqref{eq:newA}.
These generalizations require dislocation character dependent generalizations of many other parts of the model, as described in detail below.

In Section \ref{sec:results} we first present simulation results for flyer plate impact experiments of copper and aluminum.
In discussing the temperature dependence of the elastic precursor decay  of aluminum, we connect our work to previous results.
We then proceed with a sensitivity study at the example of copper, focusing on dislocation drag, character dependence, and dislocation density evolution, aiming at disentangling how these model ingredients influence the elastic precursor decay.
The same qualitative conclusions are applicable to aluminum and likely many other face centered cubic (fcc) metals.
Even though we include the same rough estimates of the temperature dependence of the elastic constants as Refs. \cite{Lloyd:2014JMPS,Austin:2018}, our results are clear enough to conclude in Section \ref{sec:comparison} that the increase in elastic precursor amplitude in aluminum at elevated temperatures can only be understood if dislocation density evolution behavior is temperature dependent as well\footnote{
Zuanetti et al. \cite{Zuanetti:2021} recently introduced a temperature dependent dislocation density saturation function for polycrystalline aluminum.
Its generalization to single crystals as well as a fully temperature dependent model for dislocation multiplication and nucleation rates has yet to be studied and this is left for future work.}.

\section{Summary of the single crystal plasticity model and its implementation}
\label{sec:modelsummary}
%%%%%%%%%%%%%%%%%%%%%%%%%%%%%%%%%%%%%%%%%%%%%%%

Following \cite{Luscher:2016,Mayeur:2016}, we consider a system of coupled equations consisting of continuum solid dynamics --- termed \emph{deformation momentum balance}  (DMB) --- and \emph{continuum dislocation transport} (CDT).
An additional set of compatibility relations, termed \emph{dislocation deformation compatibility} (DDC) is also required.

Let us start by reviewing the DMB subproblem:
We aim at keeping it concise and refer to the original references above for additional details, while pointing out the differences resp. generalizations used in the present work.
The deformation gradient tensor is defined as $\mat{F}\coleq\left(\vec{\nabla}_0\otimes \vec{x}(\vec{X},t)\right)^\txt{T}$ where every particle $\vec{X}$ in the undeformed configuration is mapped to a corresponding point $\vec{x}(\vec{X},t)$ in the deformed state and $\vec{\nabla}_0$ denotes the Lagrangian gradient operator.
Furthermore, we decompose $\mat{F}=\mat{F}_\txt{e}\cdot\mat{F}_\txt{p}$ into thermo-elastic lattice deformation $\mat{F}_\txt{e}$ and plastic deformation $\mat{F}_\txt{p}$ due to dislocation motion.
Momentum conservation in Cartesian coordinates (see \cite{Blaschke:2017tpfe} for a generalization to arbitrary coordinates) reads
\begin{align}
\vec{\nabla}_0\cdot \mat{P} = \rho_0 \ddot{\vec{u}}
\,,
\end{align}
where $\mat{P} $ denotes the nominal stress tensor, $\rho_0$ is the material's initial mass density, $\vec{u}(X,t)=\vec{x}(\vec{X},t) - \vec{X}$ is the displacement vector, and $\ddot{\vec{u}}$ denotes the second time derivative of the displacement, i.e. the Lagrange acceleration.

We also assume a field of dislocation density distributed among $N_s$ slip systems, each with positive and negative polarity (corresponding to dislocations of positive/negative orientation).
In general, these may have any character $\vartheta$, but we consider a discrete ``binning''
with $N_\txt{c}$ discrete values amenable to computational implementation.
Hence, we denote by $\varrho_\pm^\a(\vth)$ the dislocation density of $\pm$ polarity and character $\vth$ on slip system $\a$.

Every slip system is uniquely defined by its Burgers vector $\vec{b}$ (i.e. the direction of slip) and its slip plane normal $\vec{n}_0$ (i.e. $\vec{b}_0\cdot\vec{n}_0=0$ where $\vec{b}_0\coleq\vec{b}/b$, and $b=\abs{\vec{b}}$).
A dislocation can have any line sense direction $\vec{t}$ within the slip plane, i.e. $\vec{t}\cdot\vec{n}_0=0$, and if $\vec{t}$ is (anti-)parallel to $\vec{b}$ we call it a screw dislocation whereas if $\vec{t}\cdot\vec{b}_0=0$ we call it an edge dislocation.
Thus, one may introduce an angle $\vth$ between line and slip directions such that $\cos\vth = \vec{t}\cdot\vec{b}_0$, i.e. screw dislocations have $\vth=0$ or $\vth=\pi$ and edge dislocations have $\vth=\pm\pi/2$.
We will henceforth call $0\le\vth\le\pi/2$ `positive' dislocations (with subscript $_+$), and $-\pi\le\vth\le-\pi/2$ `negative' dislocations (with subscript $_-$).
For slip systems which are $\pi$-symmetric, such as the 12 fcc slip systems, these ranges are complete.
For some other slip systems (such as the ones in bcc crystals), the additional ranges $\pi/2<\vth<\pi$ and $-\pi/2<\vth<0$ must be included as well, and called `positive' and `negative' respectively.
The significance of these `polarizations' is that dislocations of  opposite polarizations can annihilate if they get close, and furthermore a net difference between positive and negative dislocations on a slip system, also known as the geometrically necessary dislocation (GND) density, gives rise to lattice curvature.
Often times, including in Refs. \cite{Luscher:2016,Mayeur:2016}, only two dislocation characters are considered, namely pure screw and edge dislocations and $N_\txt{c}=2$, but here we keep our formalism more general.

The evolution of plastic deformation is specified using an expression of Orowan's relation:
\begin{align}
\dot{\mat{F}_\txt{p}}\mat{F}_\txt{p}^{-1} 
= \sum\limits_{\a=1}^{N_\txt{s}} b^\a \left(\varrho_+^\a(\vth) + \varrho_-^\a(\vth)\right) v^\a(\vth) \vec{b}_0^\a\otimes\vec{n}_0^\a
\,,
\end{align}
where $v^\a(\vth)$ are the signed scalar velocities of dislocations with character $\vth$.
Considering dislocations to be much longer than their Burgers vectors, we only track velocities perpendicular to $\vec{t}$, and that direction is slip system and dislocation character dependent.
Hence, the dislocation velocities are $\vec{v}^\a(\vth) = v^\a(\vth) \left(\vec{t}^\a(\vth) \times \vec{n}_0^\a\right)$.
The magnitude of this velocity is a function of the effective applied stress, i.e. $\sigma = \tau - \tau_b$ (with back stress $\tau_b$ subtracted from the resolved shear stress) projected onto the Schmid factor $\vec{b}_0\otimes\vec{n}_0$ (corrected by the current elastic deformation $\mat{F}_\txt{e}$).
 In other words, $\tau \coleq \det(\mat{F}_\txt{e})\tilde{b}_i \sigma_{ij} \tilde{n}_j$ with $\vec{\tilde{b}} = (\mat{F}_\txt{e}\cdot\vec{b}_0)$ and $\vec{\tilde{n}} = ((\mat{F}_\txt{e}^{-1})^T\cdot\vec{n}_0)$.
 In particular,
 \begin{align}
 v^\a(\vth) = \frac{\sgn{\sigma}}{\left(t_\txt{w} + t_{r}\right)\sqrt{\varrho_0^\a}}
 \,,
 \end{align}
where the mean spacing between obstacles is approximated by the inverse square root of the initial dislocation density $\varrho_0^\a$ on each slip system.
The velocity $v^\alpha$ is inverse proportional to the sum of  the time a dislocation spends ``waiting'' at a barrier, $t_\txt{w}$, and the free running time $t_r$ which is limited by dislocation drag from phonon wind, see e.g. \cite{Blaschke:2019a} and references therein.
Between obstacles, the free running velocity is $v_r^\a(\vth) = \frac{\sgn{\sigma}}{ t_{r}\sqrt{\varrho_0^\a}}$ and its determination using the drag coefficient is discussed in Section \ref{sec:dragcoeff} below.
The wait time is related to the stress-assisted activation energy $\Delta G(\tau)$ required to overcome a barrier and the magnitude of thermal fluctuations.
Following \cite{Austin:2011,Luscher:2016}, it takes the form
\begin{align}
t_\txt{w}^\a &= \frac{1}{\omega_0}\left(\exp\left[\frac{\Delta G^\a(\abs{\tau^\a-\tau_b^\a}/\tau_\txt{cr}^\a)}{\kb T}\right] - 1\right)
\,.
\end{align}
This expression depends on the (slip system dependent) resolved shear stress $\tau^\a$, the back stress $\tau_b^\a$, and the critical stress also known as slip resistance $\tau_\txt{cr}^\a=\tau_0+c_1b\mu^\alpha(\vth)\sqrt{\sum_\vth(\varrho^\alpha_+(\vth)+\varrho^\alpha_-(\vth)})$,
where $\tau_0$ is an intrinsic lattice slip resistance, $\varrho_\pm$ denotes the dislocation density per slip system and character (see below for details), and $c_1$ is a material constant.
For $\Delta G^\alpha(\tau,\vth)$ we consider here almost the same functional form as in Ref. \cite{Luscher:2016} (see also \cite{Bronkhorst:2007}), but generalized to include dislocation character dependence:
\begin{align}
\Delta G^\alpha(\tau,\vth)=g_0\mu^\alpha(\vth)\abs{\vec{b}}^3\left(1-\tau^p\right)^q
\,,
\end{align}
where $\mu^\alpha(\vth)=(v_c^\alpha(\vth))^2 \rho$ denotes a dislocation character-dependent effective shear modulus corresponding to the critical velocity $v_c$ as defined in Section \ref{sec:dragcoeff}.
Furthermore, $g_0$, $p$, and $q$ are model parameters.
Finally, $\omega_0$ denotes an attempt frequency which is related to the highest acoustic phonon frequency \cite{Hunter:2015}.

Let us now turn to the continuum dislocation transport (CDT) sub-problem:
On each slip system $\a$, the change in dislocation density $\varrho^\a$ of character $\vth$ (and given polarity) over time is dictated by the flux of dislocation density and a ``source'' term which models the creation, multiplication, and annihilation of dislocations, i.e. \cite{Luscher:2016} (see also \cite{Groma:2003,Arsenlis:2004,Yefimov:2005} and references therein)
\begin{align}
\dot{\varrho}^\a_\pm(\vth) + \vec{\nabla}_0\cdot \left(\varrho^\a_\pm(\vth)\, \mat{F}_\txt{p}^{-1} \cdot \vec{v}^\a_\pm(\vth)\right)
=s^\a(\vth)
\,,
\end{align}
where the dislocation velocity $\vec{v}^\a_\pm(\vth)$ was discussed above.
A simple model for dislocation multiplication and annihilation is given by the following source term \cite{Luscher:2016,Mayeur:2016} (but generalized here to include dislocation characters $\vth$):
\begin{align}
s^\a(\vth,x) = \left(C_M\sqrt{\varrho^\a_\txt{obs}(x)} \varrho^\a(\vth,x)
- C_A Y_e \varrho_+^\a(\vth,x)\varrho_-^\a(\vth,x)\right)\abs{v^\a(\vth,x)}
\label{eq:newsource_nonucl}
\end{align}
where $\varrho^\a(\vth,x) = \varrho^\a_+(\vth,x)+\varrho^\a_-(\vth,x)$ and the forest dislocation density of slip system $\a$ and dislocation character angle $\vth$ at spatial point $x$ is estimated by $\varrho^\a_\txt{obs}(x) = \sum_{\vth} A_{\a\b}(\vth)\varrho^\b(\vth,x)$.
Constants $C_M$ (multiplication), $C_A$ (annihilation), and $Y_e$ (dislocation capture distance) are model parameters.
In the simplest approximation of \cite{Luscher:2016,Mayeur:2016}, coefficient matrix $A_{\a\b}=1$.
A more sophisticated approach would be based on the assumption that only dislocations of slip systems other than $\a$ are obstacles, and this may be quantified by the expression
\begin{align}
A_{\a\b}(\vth) = \abs{\vec{n}_0^\a\cdot\vec{t}^\b(\vth)}
\,.\label{eq:newA}
\end{align}
This form further generalizes what was assumed in \cite{Luscher:2017,Nguyen:2021}, where an average over screw and edge dislocations was taken for $A_{\a\b}$ instead of explicitly resolving its character dependence.
Eq. \eqref{eq:newA} thus also generalizes to arbitrary dislocation character what was considered in Ref. \cite{Lloyd:2014JMPS,Lloyd:2014} for pure edge dislocations only.
In Section \ref{sec:results} below we compare the effect of a character dependent matrix $A_{\a\b}$ according to \eqnref{eq:newA} to $A_{\a\b}=$constant.
In addition to dislocation multiplication and annihilation, one also has to account for homogeneous dislocation nucleation in single crystals \cite{Lloyd:2014} (in contrast to heterogeneous nucleation which occurs in polycrystals \cite{Austin:2018}).
A simple model accounting for this effect is \cite{Lloyd:2014JMPS,Lloyd:2014}
\begin{align}
%s^\a_\txt{nucl}(\vth) &= \dot{\varrho}_{n_0} \exp\left(-G_{n_0}\frac{\left(1 - \left(\frac{\abs{\tau_\txt{eff}}}{\tau_{n_0}}\right)^{p_n}\right)^{q_{n}}}{\kb T}\Theta\left(1 - {\abs{\tau_\txt{eff}}}/{\tau_{n_0}}\right)\right)
%%%% keep q_n=1=p_n for now; don't need even more model parameters:
s^\a_\txt{nucl}(x) &= \dot{\varrho}_{n_0} \exp\left(-G_{n_0}\frac{\left(1 - {\abs{\tau^\a(x)}}/{\tau_{n_0}}\right)}{\kb T(x)}\Theta\left(1 - {\abs{\tau^\a(x)}}/{\tau_{n_0}}\right)\right)
\,, \label{eq:newsource_nucl}
\end{align}
which we add to $s^\a(\vth,x) $.
In this expression, the maximum nucleation rate $\dot{\varrho}_{n_0} $, as well as $G_{n_0}$ and $\tau_{n_0}$, are model parameters.
Furthermore, $T(x)$ is the local temperature at spatial point $x$, and $\tau^\a(x)$ is the local resolved shear stress projected onto slip system $\a$.

Concerning the dislocation deformation compatibility (DDC) sub-problem,
we note that the presence of dislocations leads to elastic lattice deformations and subsequently to internal stress fields giving rise to the back stress we alluded to above.
In general, the magnitude of the back stress depends on various details of the problem at hand, see e.g. \cite{Kuroda:2019}.
Since our main goal in this work is to study the effects of dislocation drag on a 1D uniaxial stress impact problem, the back stress is not particularly important in our present case, and we may hence use once more an approximation similar to what was used in \cite{Groma:2003,Evers:2004,Luscher:2016,Mayeur:2016} for the internal stress field at point $x$:
\begin{align}
\tau_i^\a(x,\vth) = \mu b \mathcal{L}_\varrho^2 \nabla_s \kappa^\a(x,\vth)
\,,\label{eq:backstress}
\end{align}
where $b$ is the Burgers vector magnitude, $\mu$ is the average shear modulus, and $\kappa$ denotes the ``GND field'', i.e. the difference in dislocations of positive and negative polarization per slip system and dislocation character.
Its gradient along the slip direction gives rise to a stress field, which is what we compute here.
Furthermore, this stress field is screened by distant dislocations which is why we presently approximate the characteristic length scale by the inverse square root of the initial total dislocation density, $\mathcal{L}_\varrho\approx 1/\sqrt{\varrho_0}$.
Note that in Ref. \cite{Mayeur:2016}, $\mathcal{L}_\varrho$ within \eqref{eq:backstress} was considered to be an independent model parameter and was chosen to be of the order of $1/\varrho_0$ (no square root) leading to a significantly larger back stress.
The present prefactor, however, has a better physical motivation.
In particular, it has been argued in Ref. \cite{Groma:2003} that $\mathcal{L}_\varrho$ is related to the mean dislocation spacing.
We have also studied a generalized back stress along the lines of \cite{Groma:2003}, but since the magnitude is similarly small to the present back stress in \eqref{eq:backstress} and hence has no visible influence on the impact simulations at hand, we postpone presenting this detail to future work.
%%% COMMENT: in the code, we use this currently, but in the new zaiser back stress we use the total current local disloc. density of given character.

The total internal stress tensor is derived by multiplying $\tau_i$ with the symmetrized Schmid factor and summing over all slip systems and dislocation characters \cite{Harder:1999,Evers:2004,Luscher:2016}:
\begin{align}
\mat{S}_i &= \sum_\a^{N_s}\sum_{\vth_0}^{\vth_{N_c}} \tau_i^\a(x,\vth) \frac{1}{2}\left(\vec{b}_0^\a\otimes\vec{n}_0^\a + \vec{n}_0^\a\otimes\vec{b}_0^\a\right)
\end{align}
Its projection onto the Schmid factor of slip system $\a$ finally yields the slip system dependent back stress referred to earlier above:
\begin{align}
\tau_b^\a &= \mat{S}_i : \left(\vec{b}_0^\a\otimes\vec{n}_0^\a\right)
\,,
\end{align}
where $(\mat{A}:\mat{B})\coleq \tr\left(\mat{A}^T\cdot\mat{B}\right) = A_{ij} B_{ij}$.
Note that the latter does not equal $\tau_i$ of that same slip system (summed over $\vth$) as it also includes contributions from projections of all other slip systems onto $\a$.
From our simulation results in Section \ref{sec:results} below, we see that the back stress is indeed very small compared to the applied stress for the simulation conditions we focus on in this work.
%(Though, there will of course be situations where the back stress becomes more important.)

\section{Dislocation drag}
\label{sec:dragcoeff}
%%%%%%%%%%%%%%%%%%%%%%%%%%%%%%%%%%%%%%%%%%%%%%%%%%%%%%%%%

Following \cite{Luscher:2016}, the mean free-running velocity
$\vec{v}^\alpha_r(\vth) = {v}^\alpha_r(\vth) \left(\vec{t}^\a(\vth)\times \vec{n}_0^\a\right)$
of dislocations of character $\vth$ within slip system $\alpha$ is determined by
\begin{align}
 {v}^\alpha_r &= \frac{b^\alpha}{B^\alpha({v}^\alpha_r)} \left(\tau^\alpha - \tau_\txt{b}^\alpha\right) %\Theta\left(\abs{\tau^\alpha} - \abs{\tau_\txt{b}^\alpha}\right)
 \,, \label{eq:runvelocity-B}
\end{align}
where $\tau^\alpha$ is the resolved shear stress, $\tau_\txt{b}^\alpha$ is the back stress and $B^\alpha({v}^\alpha_r)$ is a drag coefficient which depends on the dislocation velocities making this a non-linear equation in ${v}^\alpha_r$.
Note that ${v}^\alpha_r$ is to be understood as an \emph{average} velocity, not the actual velocity of a single dislocation.

Previous work often made use of the form
~\cite{Austin:2018,Luscher:2016,Zuanetti:2021}:
\begin{align}
 B({v}^\alpha_r) &= \frac{B_0}{1-(\beta^\alpha)^2}
 \,, &
 \beta^\alpha &= \frac{{v}^\alpha_r}{\cs}
 \,,\label{eq:B-Austin}
\end{align}
which we wish to compare to and hence include dislocation character dependent parameters  $B_0 = B(v=0)$ and $\cs$, the latter being a limiting velocity.
The new values for these parameters were determined from PyDislocDyn~\cite{pydislocdyn}, a numerical implementation of the drag coefficient theory outlined in a recent series of papers \cite{Blaschke:BpaperRpt,Blaschke:2018anis,Blaschke:2019fits}.

From equation \eqref{eq:runvelocity-B} it is possible to rewrite $B$ as a function of stress, and for the simple analytic form \eqref{eq:B-Austin}, this can be determined exactly:
\begin{align}
 B(\sigma)
 &
% =\frac{b^\alpha\abs{\sigma}}{\cs\left(\sqrt{(\xi^\alpha)^2+1}-\xi^\alpha\right)}
 =\frac{B_0}{2\xi\left(\sqrt{(\xi^\alpha)^2+1}-\xi^\alpha\right)}
 \,, \nn\\
 \xi &= \frac{\cs B_0}{2b^\alpha\abs{\sigma}}
 \,,\qquad
 \sigma\coleq {\tau^\alpha-\tau_\txt{b}^\alpha}
 \label{eq:BofsigmaAustin}
\end{align}
where $\alpha$ labels the slip system under consideration.

\begin{figure}[!h!t]
 \centering
 \includegraphics[width=0.5\textwidth]{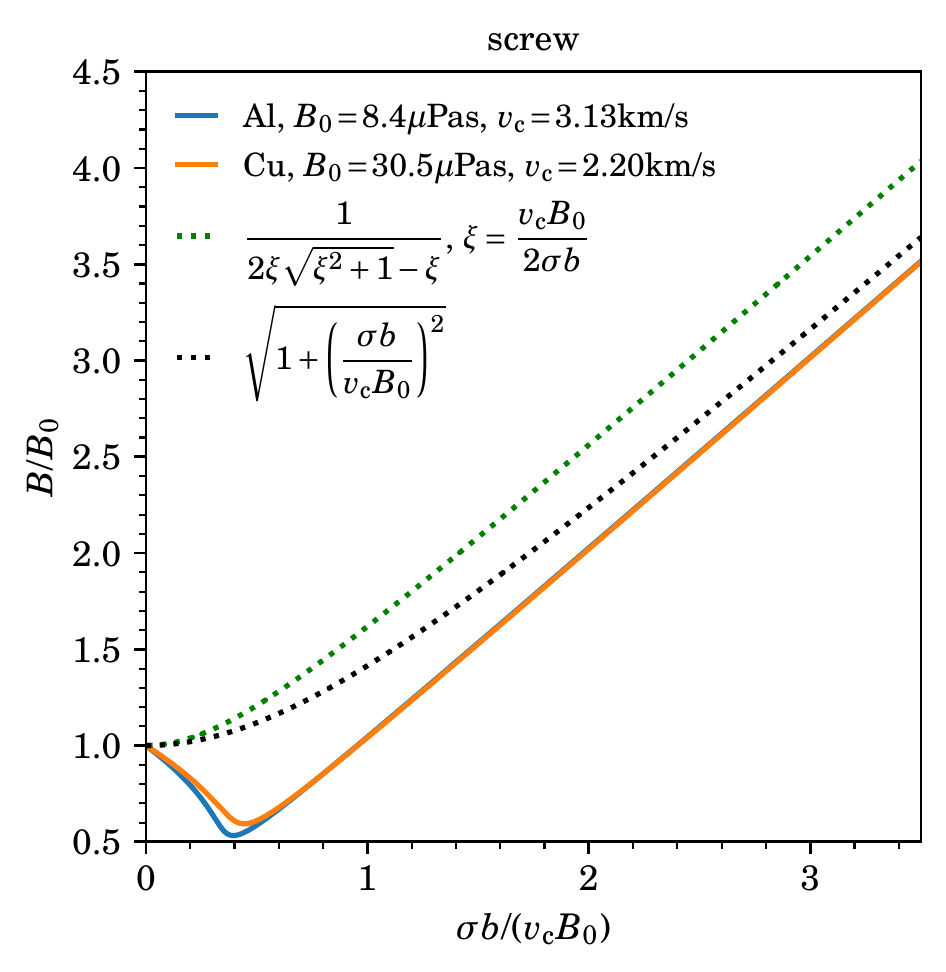}%
 \includegraphics[width=0.5\textwidth]{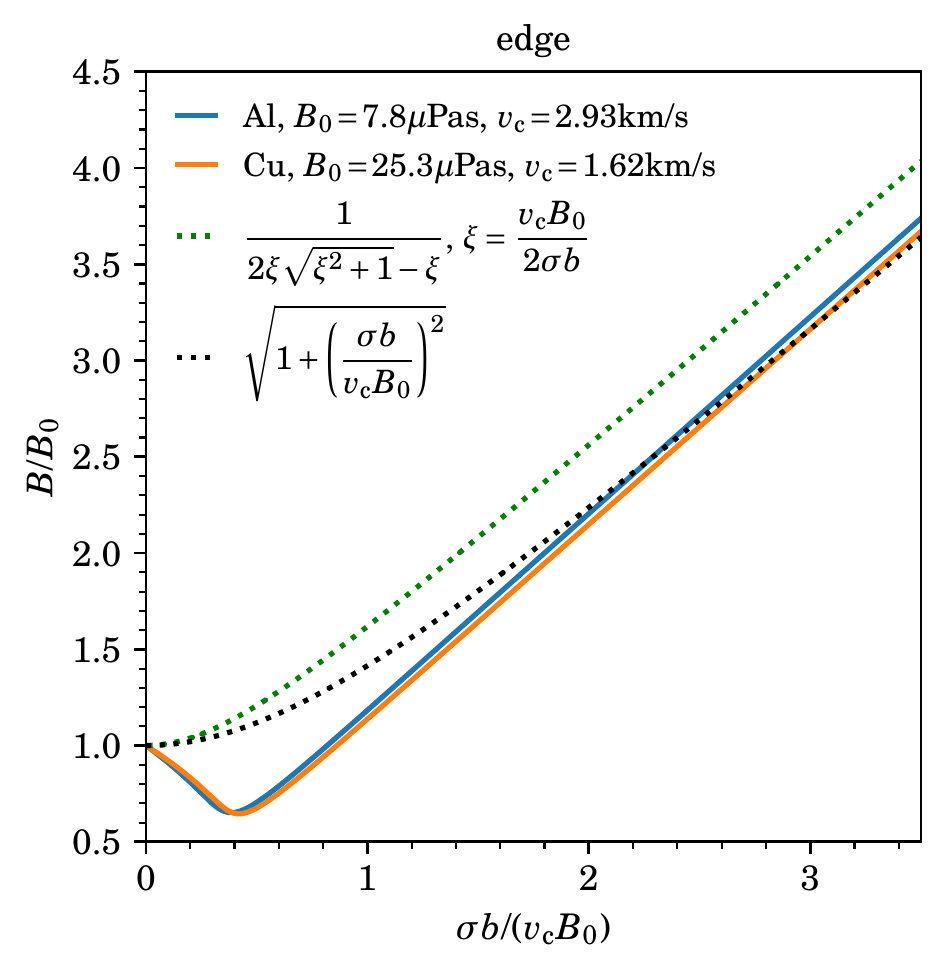}
 \caption{The dislocation drag coefficient for screw (left) and edge (right) dislocations for aluminum and copper calculated in the semi-isotropic approach \cite{Blaschke:2018anis,pydislocdyn}.
 In particular, the colored solid lines show the result of for $B(v(\sigma))$ as a function of stress using the relation $vB(v) = \sigma b$ and fitting functions for $B(v)$ along the lines of ref. \cite{Blaschke:2018anis}.
 The dashed lines show how well simple analytic functions \eqref{eq:BofsigmaAustin} (green) and \eqref{eq:Bofsigmasqrt} (black)
% with only one free parameter 
compare to the numerical result.
 Both functions reproduce the zero velocity value and tend to the correct infinite stress limit $\lim\limits_{\sigma\to\infty}B(\sigma)\approx \frac{\sigma b}{\cs}$, but \eqref{eq:Bofsigmasqrt} is closer to the true values over the whole range of stresses.
 The low stress regime can clearly be improved, but our impact simulations are not very sensitive to this regime.
 The advantage of using \eqref{eq:Bofsigmasqrt} as an approximation is that one only needs to calculate $B$ at $v=0$ since $\cs$ can be determined analytically for fcc pure screw and edge dislocations.}
\label{fig:copper-drag}
\end{figure}

More recently, it was determined that (in contrast to $B(v)$), a better approximation to $B(\sigma)$ is given by~\cite{pydislocdyn,Blaschke:2019a}
\begin{align}
 B(\sigma) &= B_0\sqrt{1+\left(\frac{b^\alpha\sigma}{\cs B_0}\right)^2}
 \,, \label{eq:Bofsigmasqrt}
\end{align}
which corresponds to
\begin{align}
 B({v}^\alpha_r) &= \frac{B_0}{\sqrt{1-(\beta^\alpha)^2}}
\,. \label{eq:Bofvsqrt}
\end{align}
Note that the best agreement of this functional form with the numerically determined $B(\sigma)$ is achieved by choosing
$B_0 \in [\min(B(v)),B(0)]$.
In Figure \ref{fig:copper-drag} we compare how well the two analytic functions \eqref{eq:BofsigmaAustin} and \eqref{eq:Bofsigmasqrt} approximate the numerically determined functional form of $B(\sigma)$ for the special cases of pure screw and pure edge dislocations, the advantage in both cases being that they can easily be implemented into a plasticity code without too much computational overhead.
Although approximation of the low stress regime of $B(\sigma)$ could clearly be improved, it was already shown in Ref. \cite{Blaschke:2019a} that plastic flow is not very sensitive to $B$ in that regime.
The same functional form can be used for mixed dislocations of arbitrary character, albeit parameters $B_0=B(v=0,\vth)$ and $\cs(\vth)$ need to be determined.
The calculation of those parameters for arbitrary dislocation character angle $\vth$ based on a first-principles theory is explained in Refs. \cite{Blaschke:2017lten,Blaschke:2018anis,Blaschke:2019fits} and a numerical implementation of this theory is given by the open source code PyDislocDyn \cite{pydislocdyn}.
As for the special cases of pure edge and pure screw in fcc metals, analytic expressions for the critical velocity $\cs$ were determined in \cite{Blaschke:2017lten} (edge) and more recently in \cite{Blaschke:2020MD} (screw):
\\
$\cs^\mathrm{fcc}(\vth=\pi/2)=\sqrt{\mathrm{min}(c',c_{44})/\rho}$ and $\cs^\mathrm{fcc}(\vth=0)=\sqrt{\frac{3c'c_{44}}{\rho(c_{44}+2c')}}$ with $c'=(c_{11}-c_{12})/2$ and $\rho$ denoting once more the material density.

In Section \ref{sec:results} below we will show how the different forms of $B$, i.e. \eqnref{eq:B-Austin}, \eqnref{eq:Bofsigmasqrt}, $B=$const., as well as different dislocation character binnings influence the elastic precursor decay within a simple 1-D shock impact problem.

\paragraph{Temperature and density dependence of dislocation drag}\ \\
%%%%%%%%%%%%%%%%%%%%%%%%%%%%%%%%%%%%%%%%%%%%%%%%%%%%%%%
%
A first rough estimate of how the dislocation drag changes with temperature and pressure was given in Ref. \cite{Blaschke:2019a} for the isotropic limit.
We generalize this estimate for anisotropic crystals by using an \emph{average} change in shear modulus, i.e. we define
\begin{align}
%\mu_0 &= \frac{1}{4}\left(c_{11} -  c_{12} + 2 c_{44}\right)
\Delta \mu(T,\rho) & = \frac{1}{4}\left(\partial_T c_{11} - \partial_T c_{12} + 2\partial_T c_{44}\right)(T-T_\mathrm{ref}) + \frac{1}{4}\left(\partial_P c_{11} - \partial_P c_{12} + 2\partial_P c_{44}\right)P(\rho,T)
\,,
\end{align}
where $T_\mathrm{ref}=300$K is the temperature at which the elastic constants and their derivatives were measured, and the current temperature $T$ and pressure $P$ are determined using the equation of state as described below in Section \ref{sec:eos}.
The according current density $\rho$ corresponding to pressure $P$ is also known at every time step in our simulation.

We therefore estimate the temperature and density (resp. pressure) dependence of $B_0(\vth)$ and $\cs(\vth)$ as:
\begin{align}
\mu_\mathrm{eff} \left(\vth,T,\rho\right) & \coleq \rho_\txt{rt} \cs^2(\vth) + \Delta \mu (T,\rho)
\,,\nn\\
B_0(\vth,T,\rho) & = B_0(\vth,\rho_\txt{rt},300)\frac{T}{300}\left(\frac{\rho}{\rho_\txt{rt}}\right)^{7/6}\sqrt{\left(1-\frac{\Delta\mu(T,\rho)}{\mu_\mathrm{eff}(\vth,T,\rho)}\right)}
\,,\nn\\
\cs(\vth,T,\rho) &= \sqrt{\frac{\mu_\mathrm{eff} \left(\vth,T,\rho\right)}{\rho}}
\,,
\end{align}
where $\rho_\txt{rt}$ denotes the reference density at ambient conditions, and hence
\begin{align}
B(\vth,T,\rho) & = B_0(\vth,T,\rho) \sqrt{1+\left(\frac{\sigma\, b^\alpha(\rho)}{\cs(\vth,T,\rho) B_0(\vth,T,\rho) }\right)^2}
\,, & b^\alpha(\rho) &= b^\alpha\frac{\rho_\txt{rt}^{1/3}}{\rho^{1/3}}
\,. \label{eq:BTrho}
\end{align}
Note that these expressions are not exact, but rather a first order approximation, where we also neglected the change in third order elastic constants with temperature $T$ and material density $\rho$.

\section{Equation of state}
\label{sec:eos}
%%%%%%%%%%%%%%%%%%%%%%%%%%%%%%%%%%%%%%%%%%%%%%%%%%
Like in Ref. \cite{Luscher:2016} we separate the Helmholtz free energy into the following contributions:
\begin{align}
\psi & = \psi_0(V) + \psi_\txt{ion}(V,T) + \psi_\txt{el}(V,T) + \psi_d(\bar{\mat{E}}_e,T) + \psi_s(\varrho_c)
\,,\nn\\
\psi_0(V) & = \frac{4 V^* B^*}{(B_1-1)^2}\left(1-(1+X)\right)\exp(-X)
\,, \qquad\qquad X = \frac{3}{2}(B_1-1)\left(\left(\frac{V}{V^*}\right)^{1/3}-1\right)
\,,\nn\\
\psi_\txt{ion}(V,T)  & = \frac{R}{M_\txt{mol}}\left(\frac{9}{8}\kb T_D+3\kb T\ln\left(1-\exp(-T_D/T)\right)-\frac{3T^4}{T_D^3}\int\limits_0^{T_D/T}\frac{z^3}{e^z-1}dz\right)
\,,\nn\\
T_D(V) & = T_{D0}\left(\frac{V}{V_0}\right)^{-\gamma^\infty}\exp\left(-a\left(\frac{V}{V_0}-1\right)-\frac{b}{2}\left(\frac{V^2}{V_0^2}-1\right)\right)
\,,\nn\\
\psi_\txt{el}(V,T) & = -\frac{1}{2}\Gamma_0T^2\left(\frac{V}{V_0}\right)^{\kappa_s}
\,,
\end{align}
where $\psi_d(\bar{\mat{E}}_e,T)=\frac{1}{2}(\bar{\mat{E}}_e)_{ij}(\bar{\mat{E}}_e)_{kl}C_{ijkl}$  is a contribution to the free energy from recoverable `deviatoric' strain and $\psi_s(\varrho_c)$ is free energy storage associated with the local state and non-local distribution of dislocation density.
Furthermore, the static lattice contribution to free energy, $\psi_0$, is based on the empirical state function developed by Ref. \cite{Vinet:1986}, where for cubic crystals the reference bulk modulus $B^*=(c_{11}+2c_{12})/3$, and its pressure derivative at ambient reference volume $V^*=1/\rho$ is $B_1 = (\partial_Pc_{11}+2\partial_Pc_{12})/3$.
For the free energy contribution from ion motion, we use the model developed by Greeff et al. in Ref. \cite{Greeff:2006}, where $T_{D0}$ is the Debye temperature at a reference volume $V_0$, $R$ is the molar gas constant, and $M_\txt{mol}$ is the molar mass.
The Debye temperature at volume $V$ is described by two model parameters $a$, $b$ describing the volume dependence of the Gr{\"u}neisen parameter:
$\gamma(V) = \gamma^\infty + a V/V_0 + b (V/V_0)^2$ with $\gamma^\infty=2/3$.
The electronic excitation free energy, $\psi_\txt{el}$ depends on the Sommerfeld coefficient $\Gamma_0$ at the reference state and $\kappa_s$ is a parameter reflecting the relationship between Sommerfeld coefficient and Fermi energy over a range of states.

\section{Results for a 1-D shock impact problem}
\label{sec:results}
%%%%%%%%%%%%%%%%%%%%%%%%%%%%%%%%%%%%%%%%%%%%%%%%%%%

In the following, we simulate classic flyer plate impact and shock load experiments for copper and aluminum.
These simulations are quasi-1D in the sense that we track the change in material deformation, dislocation density, and dislocation velocity only along the $x$-direction which is parallel to the shock wave.
Nonetheless, the full crystal anisotropy, stress tensor, and all 12 fcc slip systems are resolved at every point along $x$.
For this purpose, we use the same numerical implementation as discussed in Refs. \cite{Luscher:2016,Mayeur:2016}, but including the improvements outlined in the preceding sections with regard to drag, character dependence, and dislocation density evolution.

\begin{table}
\centering
\begin{tabular}{c|c|c|c}
 & Al & Cu & Ref.\\\hline
 $\rho$ [g/ccm], $\mu$ [GPa] & 2.7, 26.1 & 8.96, 48.3 
 & \cite{CRCHandbook} \\
$c_{11}$, $c_{12}$, $c_{44}$ [GPa] & 106.75, 60.41, 28.34 & 168.3, 121.2, 75.7 & \cite{CRCHandbook} \\
$\partial_Tc_{11}$, $\partial_Tc_{12}$, $\partial_Tc_{44}$ [MPa/K] & $-35.1$, $-6.7$, $-14.5$ & $-41.0$, $-19.8$, $-26.9$ 
& \cite{Thomas:1968}\,(Al), \cite{Luscher:2013}\,(Cu) \\
$\partial_Pc_{11}$, $\partial_Pc_{12}$, $\partial_Pc_{44}$ & 6.35, 3.45, 2.10 & 5.93, 5.05, 2.32 
& \cite{Thomas:1968}\,(Al), \cite{Luscher:2013}\,(Cu) \\
$\abs{\vec{b}}=a_l/\sqrt{2}$ [mm] & $3\times10^{-7}$ & $3\times10^{-7}$ 
& \cite{CRCHandbook} \\\hline
%%%%% EOS params:
$V_0$ [ccm/g], $T_{D0}$ [K] & $1/2.73$, 428 & $1/\rho$, 321.7 
& \cite{Jacobs:2010}\,(Al), \cite{Luscher:2016}\,(Cu) \\
$a$, $b$ & $-0.95$, $2.78^*$ & 1.259, 0.0747 
& \cite{Burakovsky:2004}\,(Al), \cite{Luscher:2016}\,(Cu) \\
$\Gamma_0$ [J/Mg], $\kappa_s$ & 50.034, $2/3$ & 11.130, 0.717 
& \cite{Burakovsky:2004}\,(Al), \cite{Luscher:2016}\,(Cu) \\\hline
%%%% disloc. drag:
$B_0$ [$\m$Pas] (screw,edge) & $(9, 8)$ & $(31, 24)$ 
& \cite{pydislocdyn} \\
$v_c$ [km/s] (screw,edge) & $(3.13, 2.93)$ & $(2.20, 1.62)$ 
& \cite{pydislocdyn}
\\
%%%% slip resistance / wait time:
$\omega_0$ [1/s], $\tau_0$[MPa], $c_1$ & $8\times10^{11}$, 20, 1/2 & $8\times10^{11}$, 20, 1/2 
& \cite{Mayeur:2016} \\
$g_0$, $p$, $q$ & 0.65, 1/3, 2/3 & 0.87, 1/3, 2/3 
& \cite{Austin:2012}\,($g_0$), \cite{Mayeur:2016}\,($p,q$) \\
%% COMMENT: \cite{Austin:2012} use p=0.5 and q=2
%%% dislocation multiplication and annihilation:
$C_M$, $Y_e$ [mm], $C_A$ & 1/30, 6$\abs{\vec{b}}$, 2 & 1/30, 6$\abs{\vec{b}}$, 2 
& \cite{Luscher:2016} \\
%%%% dislocation nucleation:
$\dot{\varrho}_{n_0}$, $G_{n_0}$ [J], $\tau_{n_0}$  & $10^{13}$, $0.12\mu\abs{\vec{b}}^3$%=8.4564\times10^{-17}$
, 115 & $5\!\!\times\!\!10^{11}$, $0.12\mu\abs{\vec{b}}^3$%=1.32084\times10^{-16}$
, 115 \\ %% where mu is the average shear modulus
%$p_n$, $q_n$ & 1, 1 & 1, 1 \\
\hline
%%%%%% initial conditions
$T$ [K] & 300 & 300 \\
$v_\txt{impact}$ [m/s] & 300 & 300 \\
$\varrho_\txt{disloc}$ [mm$^{-2}$] & $3\!\times\!10^6$  & $10^6$ \\
thickness [mm] & 0.5 & 2.0
\end{tabular}
\caption{We list the input data and model parameters used in our simulations.
Sources for these data are listed in the last column; the Burgers vector magnitude $\abs{\vec{b}}$ in fcc slip systems is related to the lattice constant $a_l$ via $\abs{\vec{b}}=a_l/\sqrt{2}$ and was rounded to 1 \r{A} accuracy for both materials.
Sources for most of these data are Refs. \cite{CRCHandbook,Thomas:1968,Burakovsky:2004,Jacobs:2010,Greeff:2006,Bronkhorst:2007,Austin:2012,Luscher:2016,Mayeur:2016}.
The model parameters for dislocation nucleation in aluminum were chosen such that the experimental room temperature results quoted in Refs. \cite{Austin:2018,Zuanetti:2021} roughly match.
(We cannot expect a perfect match since we simulate a single crystal whereas \cite{Austin:2018,Zuanetti:2021} studied a polycrystal.)
\\
\footnotesize{$^*$ These numbers for $a$, $b$ were derived from the model given in Ref. \cite{Burakovsky:2004},
$\gamma = 0.5 + \g_1 (V/V_0)^{1/3} + \g_2 (V/V_0)^q$ with model parameters $\g_1$, $\g_2$, and $q$ for aluminum, by using a least squares fit in the range $0.95 \le V_0/V \le 1.3$.
Note that our present simulations are not very sensitive to $a$ and $b$.}
}
\label{tab:inputdata}
\end{table}

\begin{figure}[!h!t]
\centering
\figtitle{Copper, 300K}
\includegraphics[trim=0 0.2cm 2.57cm 0,clip,width=0.41\textwidth]{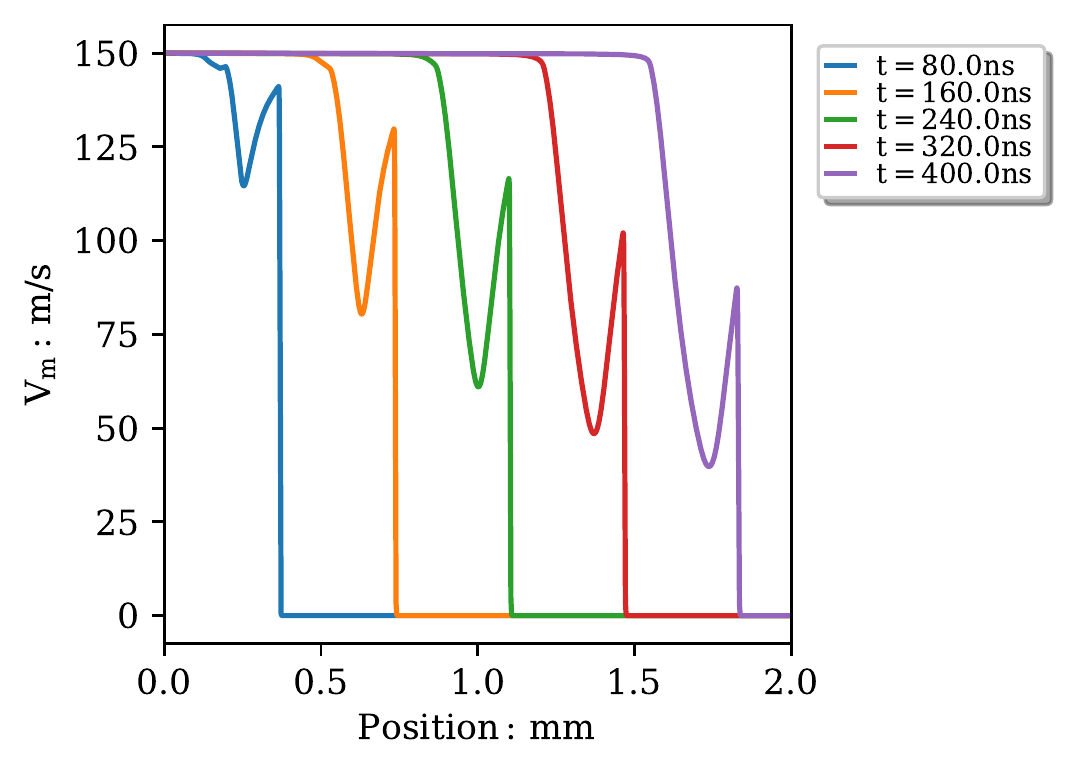}%
\includegraphics[trim=0 0.2cm 0 0,clip,width=0.59\textwidth]{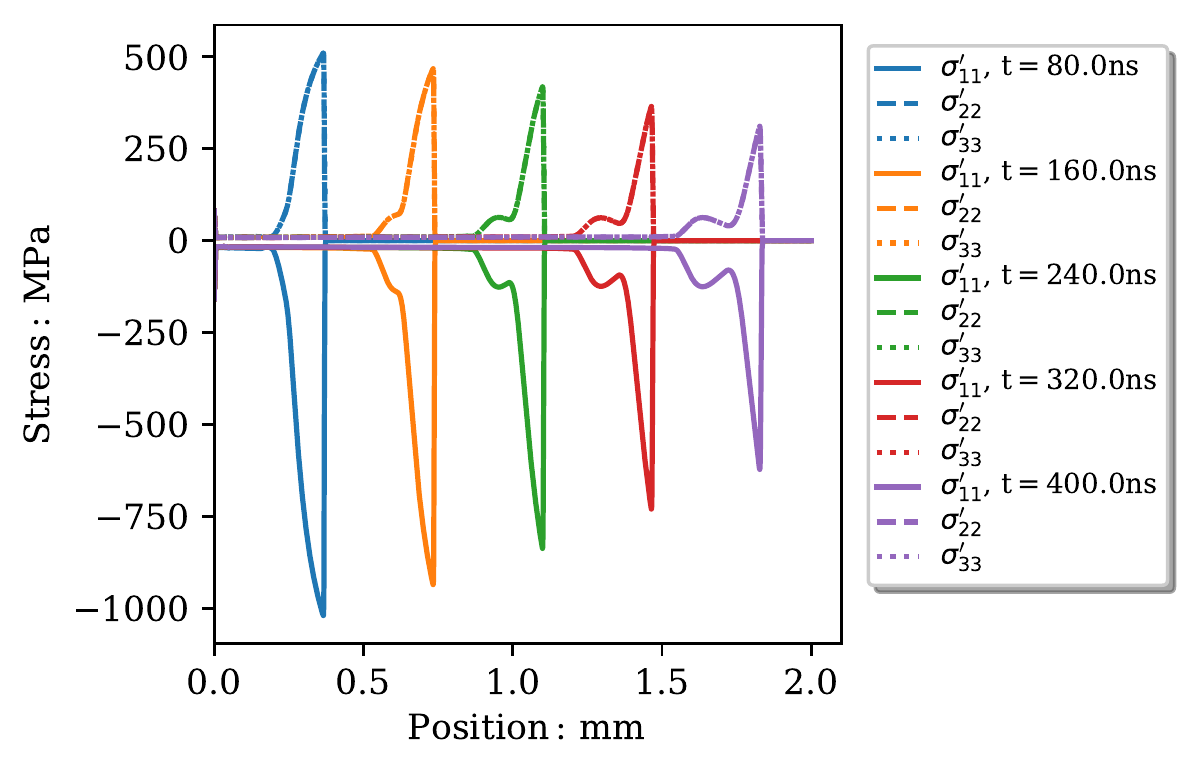}
\includegraphics[trim=0 0.2cm 0 0,clip,width=0.47\textwidth]{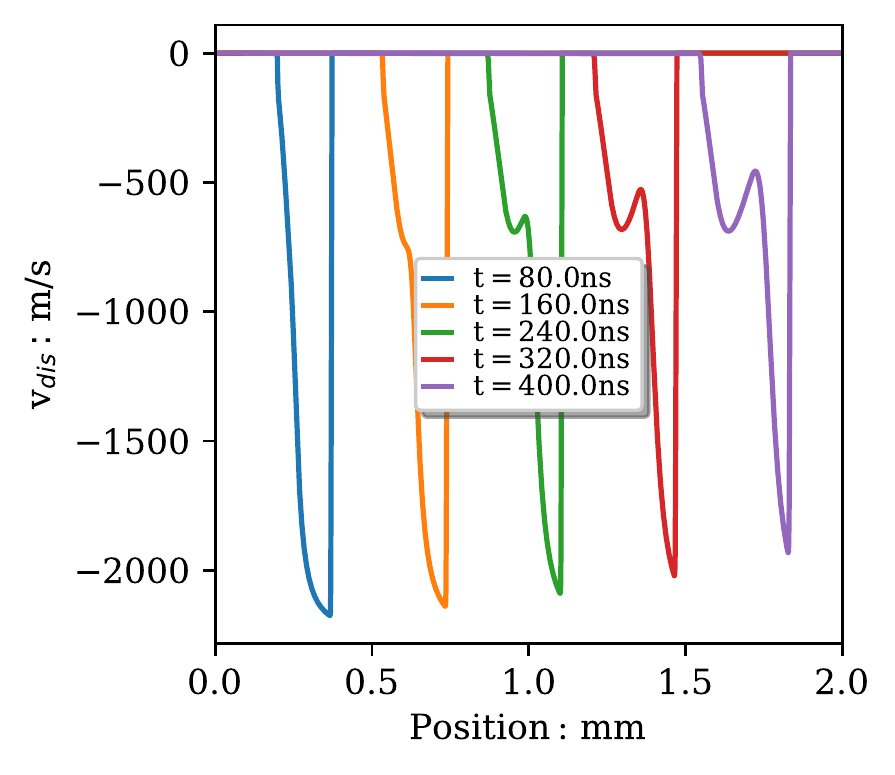}%
\includegraphics[trim=0 0.2cm 0 0,clip,width=0.44\textwidth]{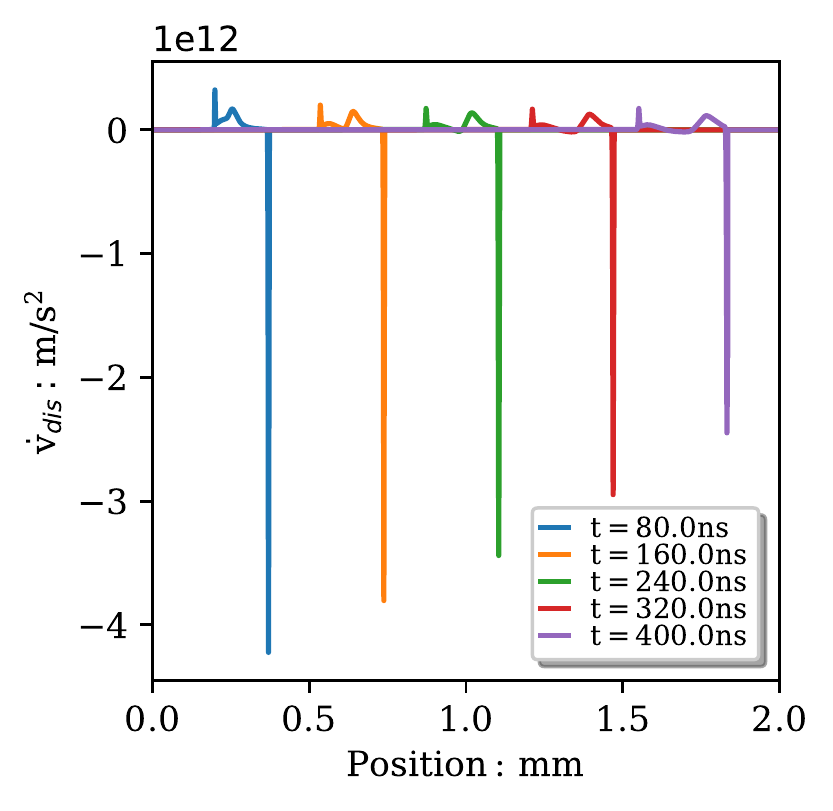}\phantom{somespace}
\caption{We show various properties of a 1D flyer plate impact simulation with one single crystal grain of copper with reference simulation parameters listed in Table \ref{tab:inputdata}.
Here, the new character dependent models for dislocation drag and dislocation density evolution described above were used.
The elastic precursor decay is clearly seen on the top left plot, which shows the material velocity parallel to the impact direction as a function of coordinate $x$ for a number of different simulation times.
On the top right, we show 
%the temperature evolution as a function of $x$; the simulation was started at room temperature, resp. 300K.
the deviatoric stress $\sigma'_{ij}=\sigma_{ij}-\frac13\delta_{ij}\tr\sigma$ and the free surface velocity at $x=2$mm.
The bottom left shows the magnitude of the pure screw dislocation velocity
%as a function of $x$ for different time stamps 
for one of the 12 slip systems.
% within the simulation.
We see that the impact velocity of 300 m/s at the fairly low dislocation density of $\varrho_\txt{disloc} =10^6$ mm$^{-2}$ drives the dislocations close to their critical velocity $v_c$.
% (which in fcc metals is lower for edge than for screw dislocations \cite{Blaschke:2020MD}).
In the bottom right, we show the acceleration these dislocations experience.
% at the same time-snapshots as on the left.
}
\label{fig:reference}
\end{figure}

\begin{figure}[!h!t]
	\centering
	\figtitle{Copper, compare to experiment}\\
	\includegraphics[trim=0 0.3cm 0 0,clip,width=0.7\textwidth]{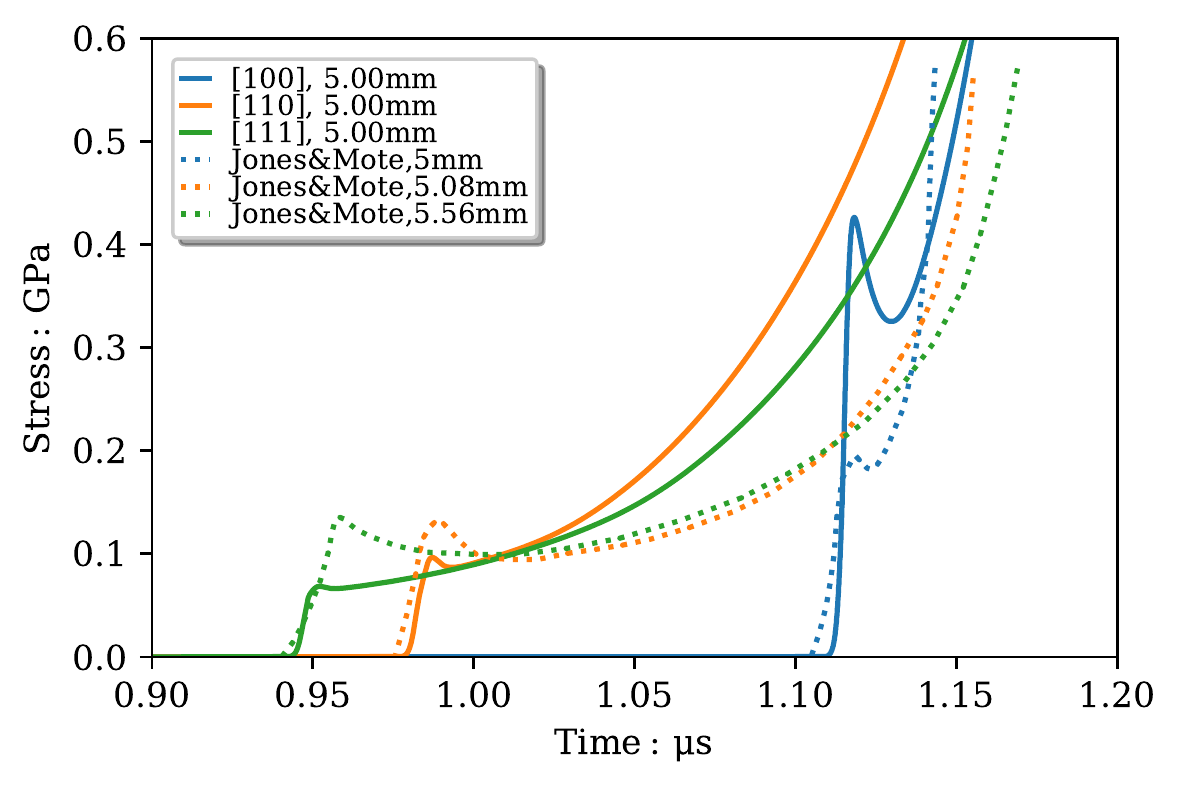}
	\caption{Our simulation results (solid lines) are compared to the experiment conducted in Ref. \cite[Fig. 3]{Jones:1969} (dotted lines), namely the deviatoric stress $\sigma_{xx}$ as a function of time for a 5mm copper target shock-loaded at 0.5 GPa along three different directions with respect to its single crystal orientation:
	along [100] (blue), [110] (orange), and [111] (green).
%		The solid lines show the results of our simulations while the dotted lines show the experimental data of Ref. \cite[Fig. 3]{Jones:1969}.
		In the latter case, the data was translated in time, which is admissible because the time-axis in that paper starts when `the first disturbance is detected' whereas in our simulation time t=0 represents the time of the initial shock load.
	}
	\label{fig:jones}
\end{figure}

\begin{figure}[!h!t]
\centering
\figtitle{Copper, compare to Taylor's simple model}
\includegraphics[trim=0 0.3cm 0 0,clip,width=0.7\textwidth]{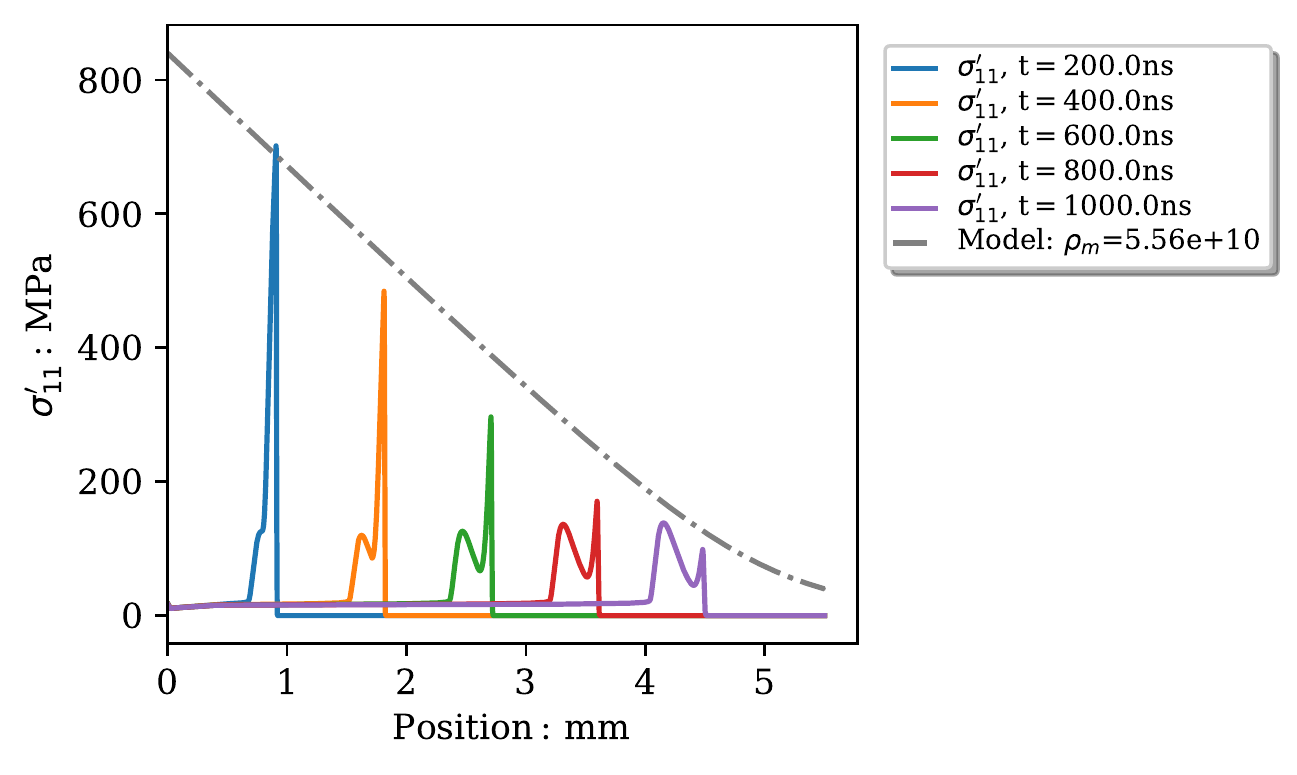}
%\vspace{-0.3cm}
\caption{We show that the peaks of deviatoric stress component $\sigma'_{xx}$ for the 5mm simulation with impact direction [100] almost follow a Taylor-like model for precursor decay; `Taylor-like' because the parameters of the  1965 Taylor model \cite{Taylor:1965} are generalized to anisotropic single crystals as described in the main text, and the new drag coefficient \eqref{eq:Bofvsqrt} is used for the strain rate that enters this model via Orowan's relation.
The initial stress was read from the simulation at an early time (t=50ns), and $\rho_m$ is a fitting parameter in this case whose value is more than 4 orders of magnitude higher than the initial dislocation density.
%Deviations between the simple model and the full simulation are expected, as the Taylor model does not capture any dislocation density evolution.
Since the Taylor model does not capture any dislocation density evolution, deviations to the full simulation are expected.
}
\label{fig:Xmodel}
\end{figure}

%%%%%%%%%%%%%%%%%%%%%%%%%%%%%%%%%%%%%%%%%%%%%%%

\subsection{Simulation results for copper}
\label{sec:copper}
%%%%%%%%%%%%%%%%%%%%%%%%%%%%%%

%Using the same numerical implementation as discussed in Ref. \cite{Luscher:2016,Mayeur:2016}, but including the improvements outline in the preceding sections with regard to drag, character dependence, and dislocation density evolution, we simulate a classic flyer plate impact experiment.
%In particular, a flyer plate is traveling
We presently consider a flyer plate traveling 
at velocity $v_\txt{impact}$ until it impacts a single-crystal copper target along the [100] lattice vector.
Assuming that the target diameter is large compared to its thickness, we may model this as a 1-dimensional plane wave.
We then also study different alignments of the single crystal with respect to the impact direction, as discussed below.
Experimental data (i.e. stress-time wave profiles and stress-strain rate curves) on single crystal copper can be found e.g. in \cite{Jones:1969,Dusek:1976}.
%%% Comment: Jones:1969 shows stress vs time (using a quartz gauge) instead of strain vs time; and Dusek:1976 conduct split Hopkinson bar experiments showing stress-strain and stress-strain rate curves but no precursor decay
%%%
Table \ref{tab:inputdata} lists the (reference) input data used in our aluminum and copper simulations.
In section \ref{sec:sensitivity} below, we conduct a sensitivity study with regard to the implementation of the drag coefficient, dislocation character dependence, and dislocation density evolution parameters.
%, but first we present some results using the `reference' data of Table \ref{tab:inputdata}.

In Figure \ref{fig:reference}, however,
%, \ref{fig:referencedisloc}, and \ref{fig:referencestress} 
we first show simulation results for copper using the `reference' parameters listed in Table \ref{tab:inputdata}.
The impact direction is parallel to one of the (cubic) crystal axes, denoted here by $x$.
%the material velocity parallel to the impact direction at different time stamps as a function of position $x$ for copper.
The results shown in this figure were computed with $N_c=2$, i.e. considering pure screw and edge dislocations, but no mixed dislocations.
In particular, we show the material velocity parallel to the impact direction in the top left of Fig. \ref{fig:reference}.
The elastic precursor decay is clearly visible in this plot.
On the top right of that same figure, we show 
%the temperature distribution along $x$ at different time stamps.
%The bottom left figure shows 
the diagonal components of the deviatoric stress (i.e. the stress with pressure subtracted).
% and the bottom right shows the free surface velocity at $x=2$mm as a function of time.
Our simulations reveal that the back stress in our case is three orders of magnitude smaller than the resolved shear stress, and hence is almost negligible within our present study.
In the bottom left of Figure \ref{fig:reference} we also show the velocity magnitude for pure screw dislocations within one of the 12 fcc slip systems.
We see that especially at early times (and close to the impacter), dislocations reach speeds close to their critical velocities, and our simulations reveal the same is true for edge dislocations (i.e. 2.20km/s for screw and 1.62 km/s for edge in fcc copper, see Table \ref{tab:inputdata}).
These high dislocation velocities are achieved because the initial dislocation density is relatively low, i.e. increasing the dislocation density by a significant amount will yield much lower dislocation velocities.
%%% point out that precursor is not seen then and also that B becomes unimportant
It is thus the high impact and low dislocation density regime where the high velocity behavior of drag coefficient $B$ becomes important.
From the bottom right of Figure \ref{fig:reference} we see that typical peak acceleration values for dislocations within this type of simulation is a few $10^{12}$m/s$^2$.
In Ref. \cite{Blaschke:2020acc} it was shown, that (at least for an fcc pure screw dislocation) accelerations of this order of magnitude leads to only small deviations in the dislocation displacement gradient field compared to its steady-state counter part.
Thus, the approximation \eqref{eq:Bofvsqrt} for dislocation drag, which we are using throughout this work and which is based on the steady-state description of dislocation motion, can be expected to be sufficiently accurate for our present purposes.
%
%Figure \ref{fig:referencestress} finally shows the resolved shear stress (left) and back stress (right) of the same slip system as Figure \ref{fig:referencedisloc}.
%Clearly, the back stress is almost negligible except very close to the impacter within our present study.
%Furthermore
%%%

%We close this subsection 
We proceed
by presenting results of an additional simulation that we compare to the experimental results of Ref. \cite{Jones:1969}.
In order to achieve (rough) agreement with that experiment, we change the following:
We simulate a 5mm target (instead of 2mm) and instead of imposing a fixed velocity at the left boundary, we now impose a force simulating a 0.5 GPa shock load (as in the experiment of \cite{Jones:1969}).
By enlarging our simulation to 5.6mm, we can simulate having a stress `gauge' at the other side of the target, and thus can plot the deviatoric stress component $\sigma_{xx}$ at $x=5$mm as a function of time within Figure \ref{fig:jones}.
The additional 0.6mm thickness were chosen so as to avoid having a wave, which is reflected from the free surface, appear in Fig. \ref{fig:jones} within the shown time interval.
Since we already established the negligibility of the backstress within
%Figure \ref{fig:referencestress}
our simulations, and because its present implementation led to numerical instabilities near the impacter at the reduced dislocation density used within this simulation, we have neglected the backstress entirely in this simulation.
Results are shown for three different crystal orientations with respect to the impact direction, i.e. parallel to [100], [110], and [111].
Despite the small differences in precursor height compared to \cite{Jones:1969} (Fig. 3 in that paper), the trends agree.
In particular, the precursor amplitude is much larger for the [100] orientation than for the other two and the subsequent rise in stress is also much steeper, with [111] having the slowest rise in stress (and reaching 0.6 GPa after $\sim0.3\mu$s in agreement with \cite{Jones:1969}).
In \cite{Jones:1969}, time was translated such that $t=0$ corresponded in all three cases to the `first detectable disturbance'.
We choose here not to translate the figures in time (i.e. the copper sample is shock loaded at time $t=0$) in order to additionally emphasize that the precursor wave arrives at the 5mm `boundary' sooner for the [111] orientation than e.g. the [100] orientation; this is expected and in agreement with the longitudinal sound speeds in copper along those crystal directions as elaborated in more detail within the discussion of Figure \ref{fig:orientation} below.
In order to reduce the height of the minimum and maximum of the precursor amplitude within the [100] orientation, one could further enhance dislocation multiplication and/or nucleation (see our sensitivity study in Section \ref{sec:sensitivity} below), but this would equally affect the other orientations which already match the experimental results of \cite{Jones:1969} fairly well.
Better agreement across all three orientations could potentially be achieved with a more sophisticated dislocation density evolution model (which we leave for future work).
We also note, that the present 5mm simulation is very sensitive to the initial dislocation density as well as the dislocation density multiplication and nucleation parameters.

%%%%%%%% ALUMINUM PLOTS: %%%%%%%%%%

\begin{figure}[!h!b]
	\centering
	\figtitle{Aluminum: 300K vs 933K}
	\includegraphics[trim=0 0 2.4cm 0,clip,width=0.437\textwidth]{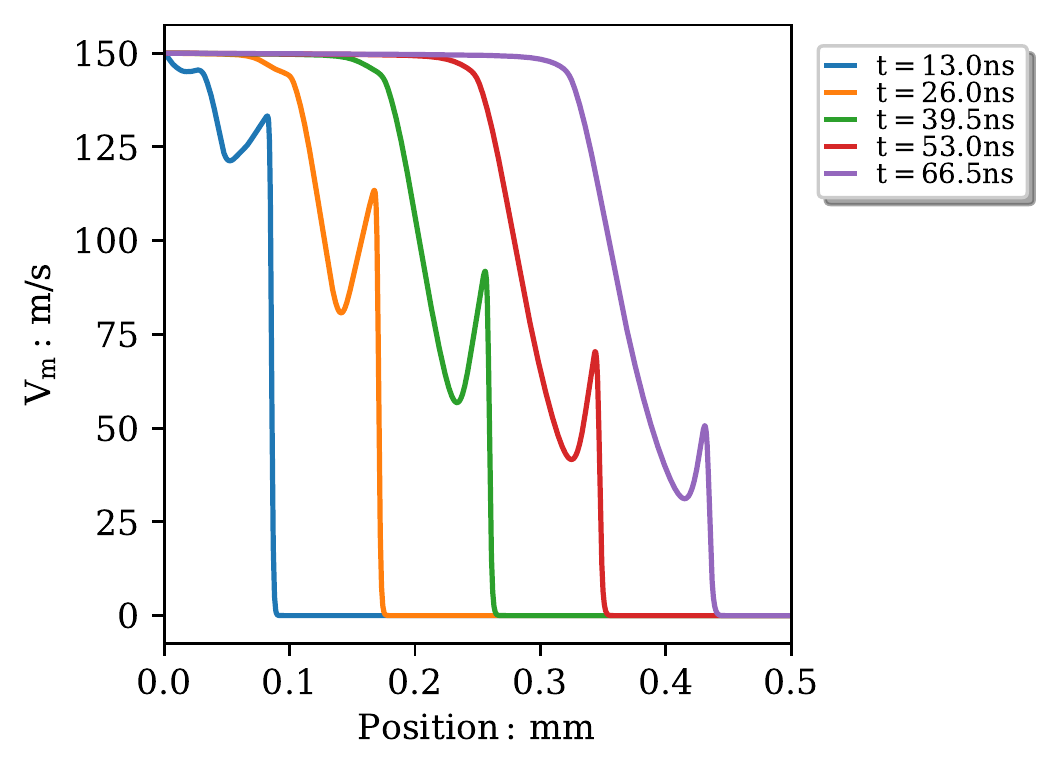}%
	\includegraphics[width=0.563\textwidth]{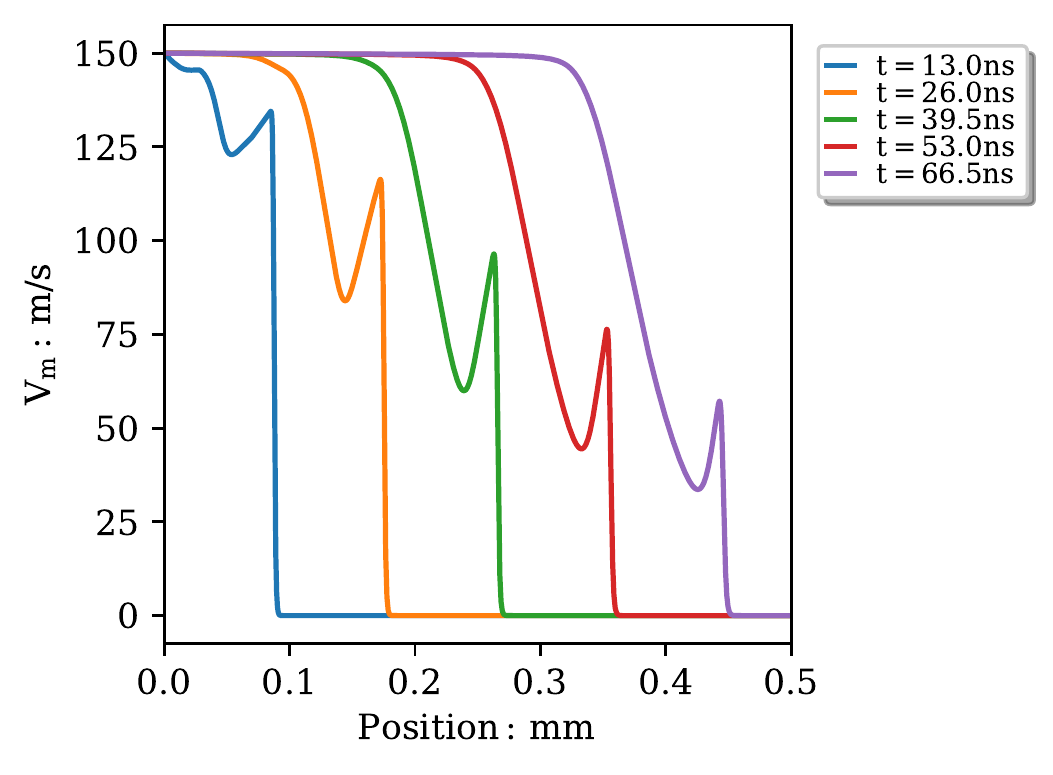}
	\includegraphics[trim=0 0.3cm 0 0,clip,width=0.45\textwidth]{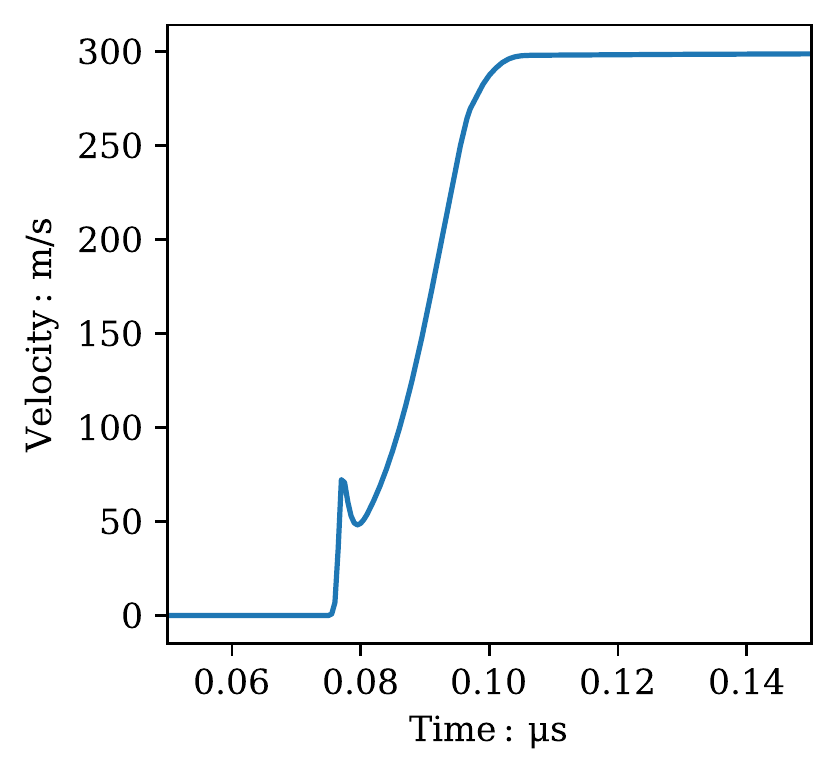}%
	\includegraphics[trim=0 0.3cm 0 0,clip,width=0.45\textwidth]{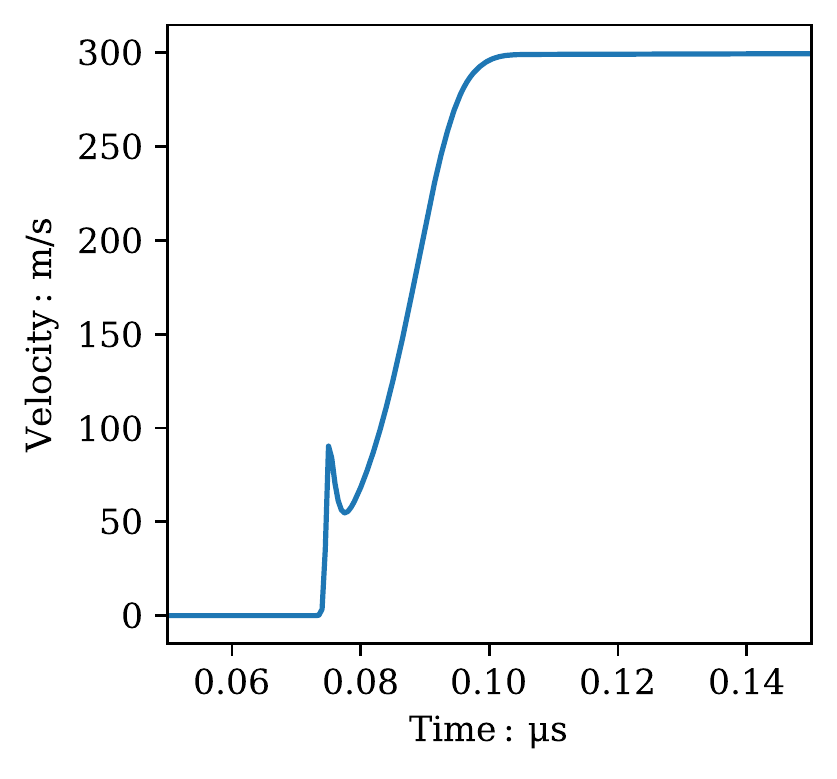}\phantom{somespace}
	\caption{We show the elastic precursor decay in a 1D flyer plate impact simulation with one single crystal grain of aluminum.
		We ran the simulations with an impact velocity of 300m/s and two dislocation characters with dislocation drag according to \eqref{eq:BTrho}.
		The top row shows simulations at $T=300$K (left) and at 933K (right) with the same density evolution parameters.
		The bottom row compares the free surface velocity for both simulations.
	}
	\label{fig:Tdep_Al1}
\end{figure}

Taylor \cite{Taylor:1965} put forward a very simple model capturing the main trend of elastic precursor decay.
To show that our simulation follows this trend to a rough first order approximation, but also to highlight the differences, we compare a simple Taylor-like model to the deviatoric stress component $\sigma'_{xx}$ within the 5mm simulation with impact direction [100] in Figure \ref{fig:Xmodel}.
Taylor's precursor decay model states that $\partial_x \sigma = -2\frac{G}{\cl}\dot\varepsilon^p$, where the isotropic shear modulus $G$ and longitudinal sound speed $\cl$ need to be generalized to our present anisotropic single crystal geometry.
Noting that shear modulus $G$ enters the original model due to its relation to the limiting dislocation velocity $\ct$ in the isotropic case, we replace it here with the effective shear modulus, $c'=\frac12(c_{11}-c_{12})$, which corresponds to the limiting velocity for pure edge dislocations in fcc crystals (see Section \ref{sec:dragcoeff} above).
It is smaller than the limiting velocity of screw dislocations and also coincides with the smallest limiting velocity for any dislocation character in fcc metals \cite{Blaschke:2017lten}, and thus can be expected to dominate.
The longitudinal sound speed in the glide direction of edge dislocations, [110], is easily determined to be $\cl^{110}=4.96$km/s for copper.
The plastic strain rate $\dot\varepsilon^p$ can be related to the dislocation velocity via Orowan's relation (though neglecting the wait time associated to thermal activation in the regime of fast moving dislocations), and we use the new drag coefficient \eqref{eq:Bofvsqrt} (with \eqref{eq:runvelocity-B} to determine $v(\sigma$)) for this purpose.
Deviations between this very simplistic model and the full simulation, as shown in Figure \ref{fig:Xmodel}, are expected, since the Taylor model does not capture any dislocation density evolution.

\begin{figure}[!h!t]
	\centering
	\figtitle{Aluminum, 933K, lower $\varrho_\mathrm{disloc}$}
	\includegraphics[trim=0 0.3cm 0 0,clip,width=0.437\textwidth]{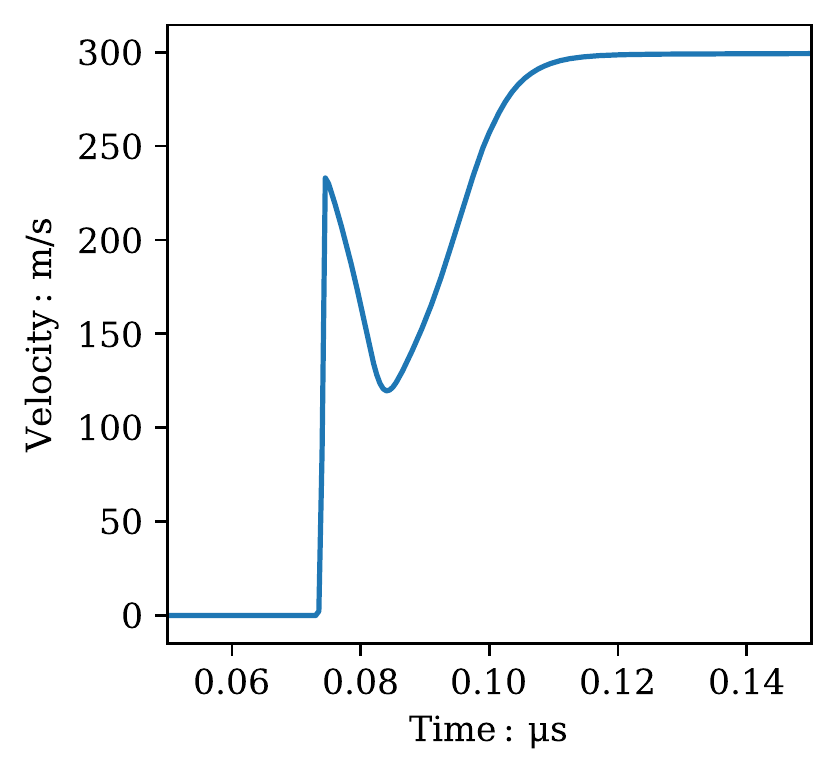}%
	\includegraphics[trim=0 0.3cm 0 0,clip,width=0.563\textwidth]{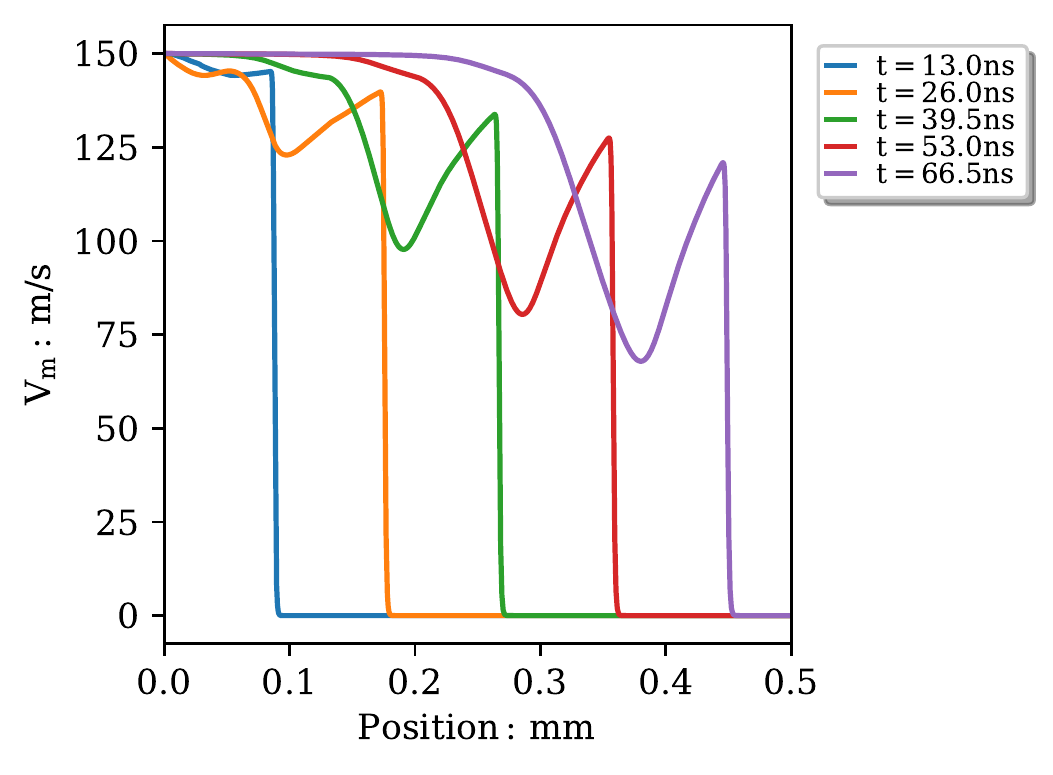}
	\caption{We show once more the elastic precursor decay results in a 1D flyer plate impact simulation with one single crystal grain of aluminum at temperature $T=933$ K with an impact velocity of 300m/s and two dislocation characters with dislocation drag according to \eqref{eq:BTrho}.
		Compared to the previous Figure \ref{fig:Tdep_Al1}, we decreased the initial dislocation density and nucleation parameter for demonstration purposes:
		In particular, $\varrho_\txt{disloc}=10^6$ [mm$^{-2}$] and $\dot{\varrho}_{n_0}=2\times10^{12}$.
		On the left we show the free surface velocity as a function of time and on the right once more the material velocity as a function of position for several different time stamps.
	}
	\label{fig:Tdep_Al2}
\end{figure}

\begin{figure}[!h!t]
	\centering
	\figtitle{Aluminum, 933K, lower dislocation density/nucleation rate}
	\includegraphics[trim=0 0.3cm 0 0,clip,width=0.5\textwidth]{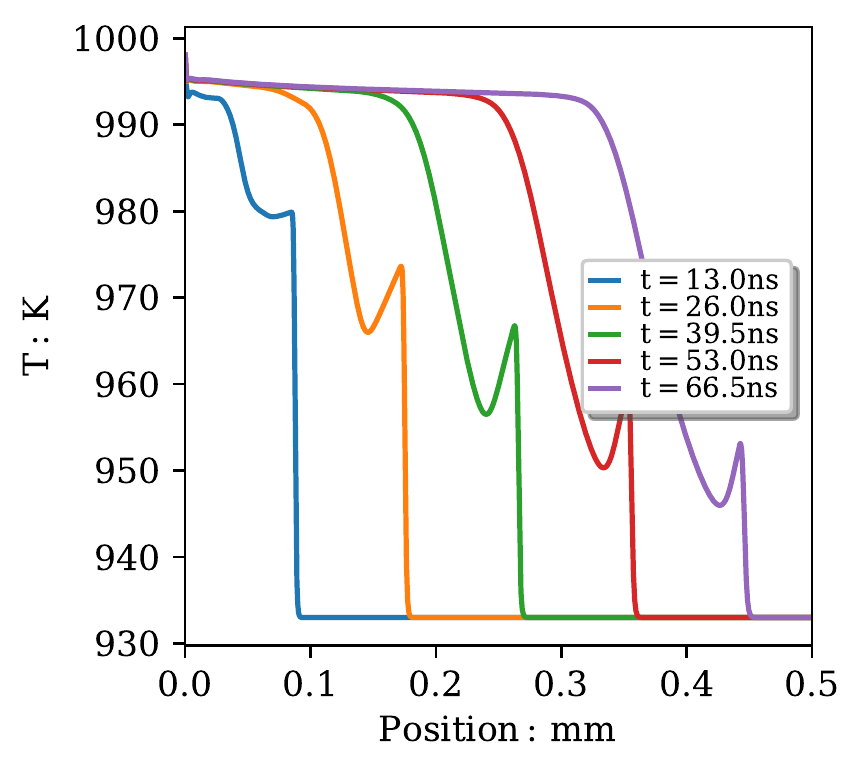}%
	\includegraphics[trim=0 0.3cm 0 0,clip,width=0.5\textwidth]{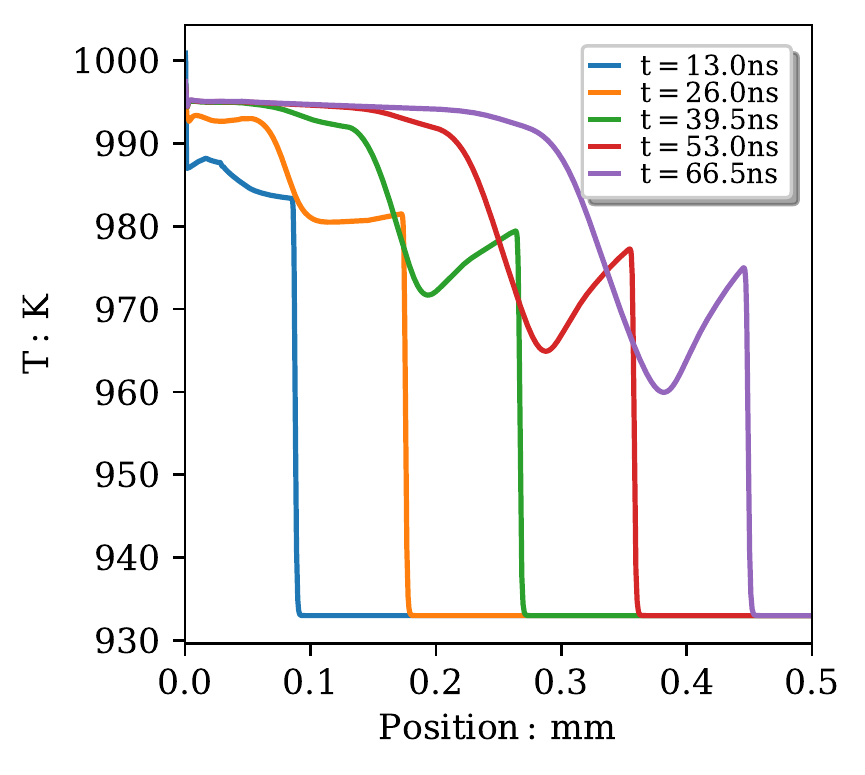}
	\caption{Comparing Figures \ref{fig:Tdep_Al1} and \ref{fig:Tdep_Al2} further, we show how the temperature profile within the 933 K simulation changes upon reducing the initial dislocation density and nucleation rate.
		In particular, on the left we had $\varrho_\txt{disloc}=3\times10^6$ [mm$^{-2}$] and $\dot{\varrho}_{n_0}=10^{13}$ whereas on the right $\varrho_\txt{disloc}=10^6$ [mm$^{-2}$] and $\dot{\varrho}_{n_0}=2\times10^{12}$.
	}
	\label{fig:Tdep_Al3}
\end{figure}

\subsection{Simulation results for aluminum}
\label{sec:aluminum}
%%%%%%%%%%%%%%%%%%%%%%%%%%%%%%

%% TODO #1: perhaps argue along the following lines: Austin:2018 showed that dislocation dynamics are important to incorporate leading to more accurate predictions of elastic precursor decay in polycrystalline aluminum compared to various phenomenological models (like SG, PTW, etc.). Therefore, stands to reason that in single crystal case the same is true, but we now need to resolve anisotropy of the crystal (which we do here).
%% TODO #2: point out that Austins preductions are better than previous models, but still far from perfect and we deduce from our study that density evolution is more important than Austin previously thought (don't think his argument of low VISAR resolution missing the peak is the true reason)
%% TODO #3: do we have access to any single crystal results for aluminum that we could compare to?
%% TODO #4: can we compare to any similar single crystal studies of elastic precursor decay, so that we may show improvement by our model?

In Ref. \cite{Austin:2018}, Austin showed that the inclusion of dislocation dynamics is crucial in order to capture elastic precursor decay properties of polycrystalline aluminum.
Phenomenological models, such as Refs. \cite{Steinberg:1980,Steinberg:1989,Follansbee:1988,PTW:2003}, cannot reproduce the elastic precursor peak at the free surface at all, as was shown by direct comparison within Fig. 9 of that paper.
It therefore stands to reason, that the inclusion of dislocation dynamics is at least as (if not more) important when studying single crystal aluminum, as we presently do.
Limiting velocities of dislocations in anisotropic crystals are slip system dependent and character dependent.
Hence, our present dislocation character dependent model generalizations are necessary in order to study single crystal aluminum, and we cannot resort to the isotropic drag model (or other isotropic model ingredients) of e.g. Ref. \cite{Austin:2018}.

Our initial conditions (apart from being a single crystal) for aluminum were chosen similarly to those in Ref. \cite{Austin:2018,Zuanetti:2021} in order to enable a qualitative comparison, see Table \ref{tab:inputdata}.
(There is no reason to expect quantitative agreement between a single crystal and a polycrystal simulation).
Classic flyer plate impact simulation results for room temperature (300K) and close to melting (933K) are shown in Figure \ref{fig:Tdep_Al1},
the only difference between the plots on the left and right being the initial temperature, i.e. initial dislocation density and density evolution parameters are exactly the same.
Even though dislocation drag is larger at the higher temperature, see \eqref{eq:BTrho}, the difference in elastic precursor decay between the two temperatures is small.
Clearly, the enhanced nucleation due to the higher temperature within  \eqnref{eq:newsource_nucl} counter acts the effect from enhanced drag.
Though the bottom row of Figure \ref{fig:Tdep_Al1} shows an increase in precursor amplitude in the free surface velocity, the effect is too small to fully explain the experimentally observed increase, see Refs. \cite{Kanel:2001,Zaretsky:2012,Gurrutxaga:2017}.

In Figure \ref{fig:Tdep_Al2}, we decreased the initial dislocation density $\varrho_\txt{disloc}$ and the maximum nucleation rate $\dot{\varrho}_{n_0}$,
%and increased dislocation annihilation and multiplication.
%The lower initial dislocation density and decreased nucleation lead to the higher spike amplitudes.
and this leads to enhanced precursor amplitudes.
In Figure \ref{fig:Tdep_Al3}, finally, we compare the temperature profiles of the two simulations at 933 K, and see that higher plasticity heating persists throughout the simulation when fewer dislocations are present.
The latter two figures clearly show that the elastic precursor amplitude is more sensitive to the dislocation density evolution than it is to the temperature dependence of dislocation drag.
Thus, the experimentally observed increase in elastic precursor amplitude can only be explained by a combination of both.

\subsection{Sensitivity study at the example of copper}
\label{sec:sensitivity}
%%%%%%%%%%%%%%%%%%%%%%%%%%%%%%

\begin{figure}[!h!t]
	\centering
	\figtitle{Cu: compare different dislocation drag models}
	\includegraphics[trim=0 0.3cm 2.585cm 0,clip,width=0.434\textwidth]{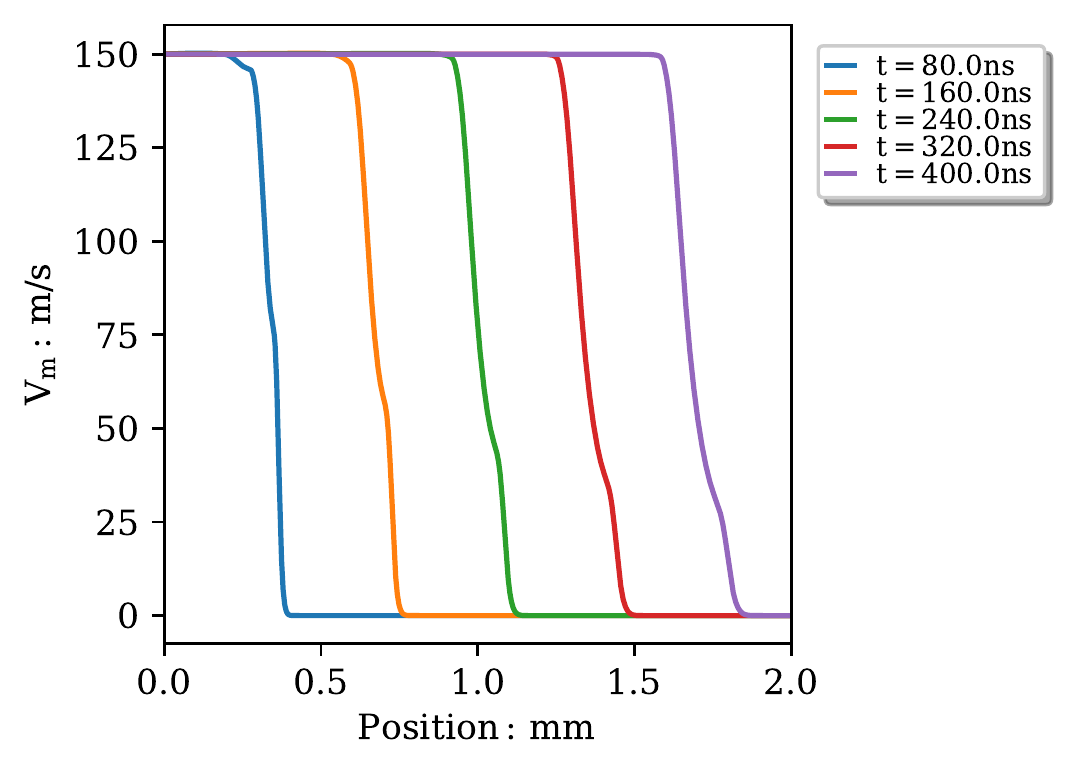}%
	\includegraphics[trim=0 0.3cm 0 0,clip,width=0.567\textwidth]{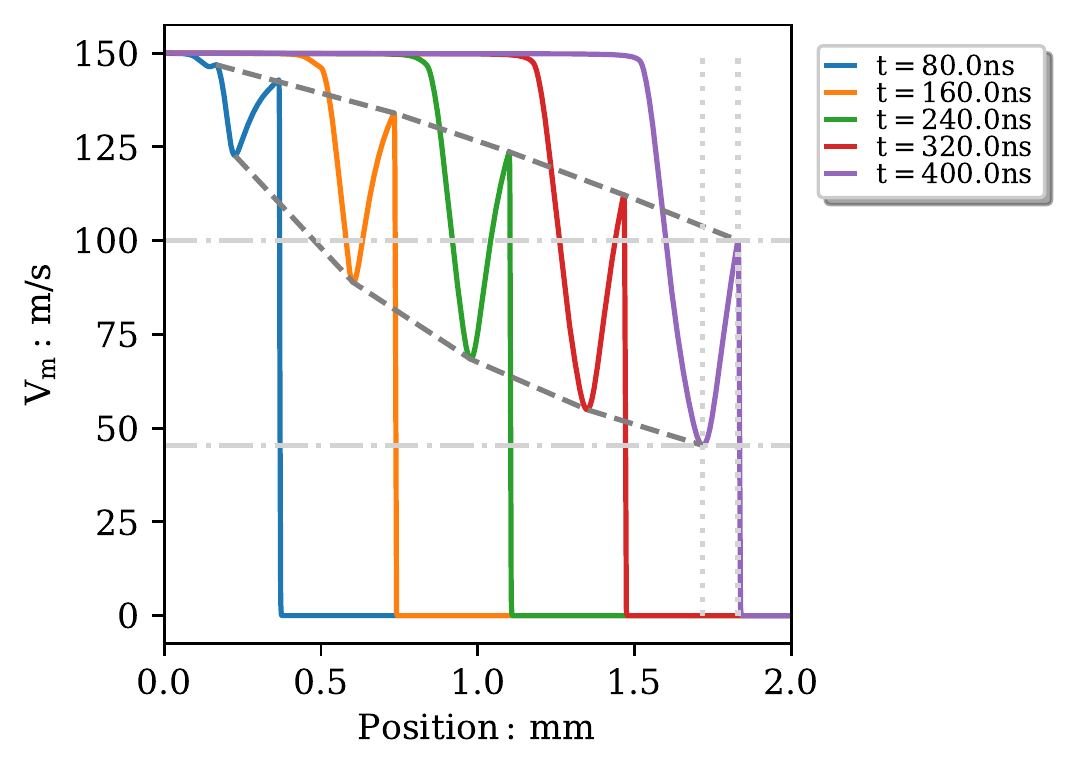}
	\includegraphics[trim=0 0.3cm 2.585cm 0,clip,width=0.434\textwidth]{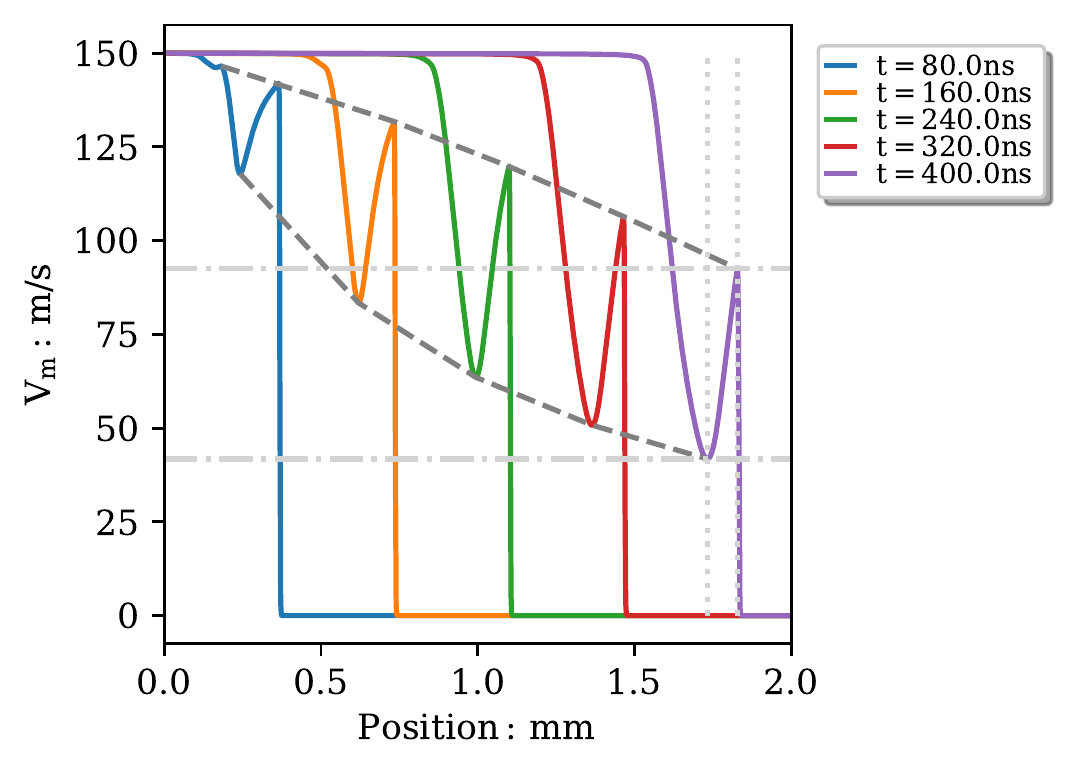}%
	\includegraphics[trim=0 0.3cm 0 0,clip,width=0.567\textwidth]{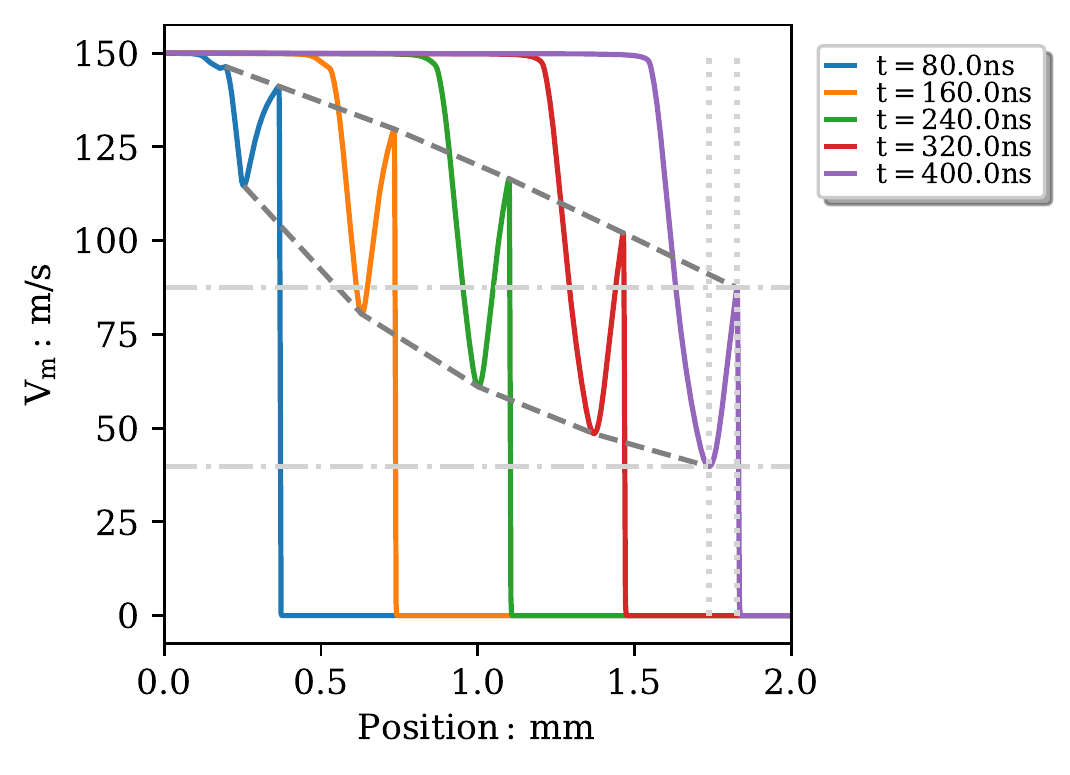}
	\caption{Results from a sensitivity study of the elastic precursor decay in a 1D flyer plate impact simulation with one single crystal grain of copper for different assumptions on dislocation drag:
	On the top left we assume a constant drag coefficient with $B_0^\txt{screw}=31\,\mu$Pa\,s and $B_0^\txt{edge}=24\,\mu$Pa\,s.
		On the top right, the empirical drag equation \eqref{eq:BofsigmaAustin} was assumed with the same zero velocity limit $B_0$ and  critical velocities $v_c$ listed in Table \ref{tab:inputdata}.
		The temperature dependence of $B_0$ and $v_c$ has been neglected in this plot: although the simulation starts at room temperature, the local temperature does change as the plastic wave moves through the crystal.
		On the bottom left, the same (temperature independent) parameters were used as on the top right, but now using the approximation \eqref{eq:Bofsigmasqrt} to the first principles derivation of drag coefficient $B$.
		On the bottom right, we considered $B_0$ and $v_c$ temperature and density dependent according to \eqref{eq:BTrho}.
%		Unless noted otherwise, reference simulation parameters listed in Table \ref{tab:inputdata} were used in all four cases.
	}
	\label{fig:dragcoeff}
\end{figure}

We proceed by changing one parameter at a time in order to see how sensitive the simulation is to the various ingredients.
In particular, we start by comparing results computed using different drag coefficient models in Figure \ref{fig:dragcoeff}:
%In Figure \ref{fig:dragcoeff}, we compare results computed using different drag coefficients:
The unphysical assumption of a constant drag coefficient yields no clear separation of elastic precursor and plastic wave, and thus no precursor decay, as shown in the top left sub-figure.
On the top right we show the precursor decay resulting from a very crude approximation to dislocation drag along the lines of \eqref{eq:B-Austin}.
Both drag models have been used repeatedly in the literature \cite{Kuksin:2008,Hansen:2013,Hunter:2015,Borodin:2015,Barton:2011,Luscher:2016,Austin:2018}.
On the bottom left we repeated the computation with a more realistic approximation to the drag coefficient, \eqnref{eq:Bofsigmasqrt}, motivated by the first-principles theory of Refs. \cite{Blaschke:BpaperRpt,Blaschke:2018anis,Blaschke:2019fits,Blaschke:2019Bpap}.
On the bottom right, we endowed the latter drag coefficient with temperature and density dependent parameters $B_0(T,\rho)$, $v_c(T,\rho)$ and Burgers vector magnitude $b(\rho)$.
%%%
The temperature and density dependence of $B_0$ and $v_c$ were estimated along the lines of \cite{Blaschke:2019a}, but generalized to the anisotropic case as described above in Section \ref{sec:dragcoeff}.
In all four cases, the impact direction is aligned with the [1,0,0] plane of the single crystal, the impact velocity is 300 m/s, and the dislocation density was assumed fairly low ($10^6\,$mm$^{-2}$) in order to make the dislocations move faster and make the simulation sensitive to the high velocity behavior of dislocation drag.
We see that the most important feature to capture is the presence, and subsequently also the accurate position, of the critical velocities (which are temperature and density dependent).
Since the empirical approximation to $B(\sigma)$, \eqnref{eq:BofsigmaAustin}, is not too far off compared to the physically motivated and more accurate approximation \eqref{eq:Bofsigmasqrt} (see Figure \ref{fig:copper-drag}), the top right and bottom left plots of Figure \ref{fig:dragcoeff} are fairly similar.
Inclusion of temperature and density dependence of the critical velocities leads to further visible changes (bottom left of Figure \ref{fig:dragcoeff}), since both temperature and density change as a function of position and time within our simulation.
%For Figure \ref{fig:dragcoeff} we started with an initial temperature of 300K, thus the differences between the last to subfigures would be larger for higher initial temperatures.
%The latter drag coefficient is the most accurate one to date, and thus should always be preferred over ad-hoc alternatives.

\begin{figure}[!h!t]
	\centering
	\figtitle{Cu: sensitivity to dislocation character resolution}
	\includegraphics[trim=0 0.3cm 2.585cm	 0,clip,width=0.434\textwidth]{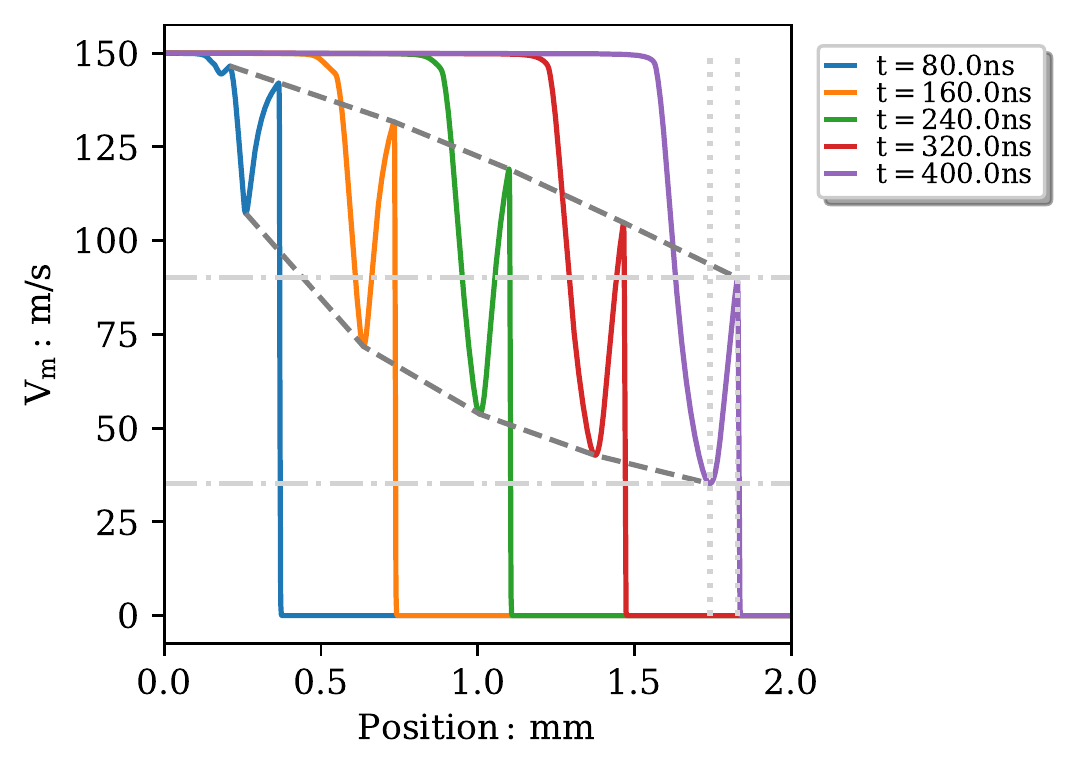}%
	\includegraphics[trim=0 0.3cm 0 0,clip,width=0.567\textwidth]{figures/copper/ref/CuRef.fig.dmb_vel_x1_wl.pdf}
	\includegraphics[trim=0 0.3cm 2.585cm 0,clip,width=0.434\textwidth]{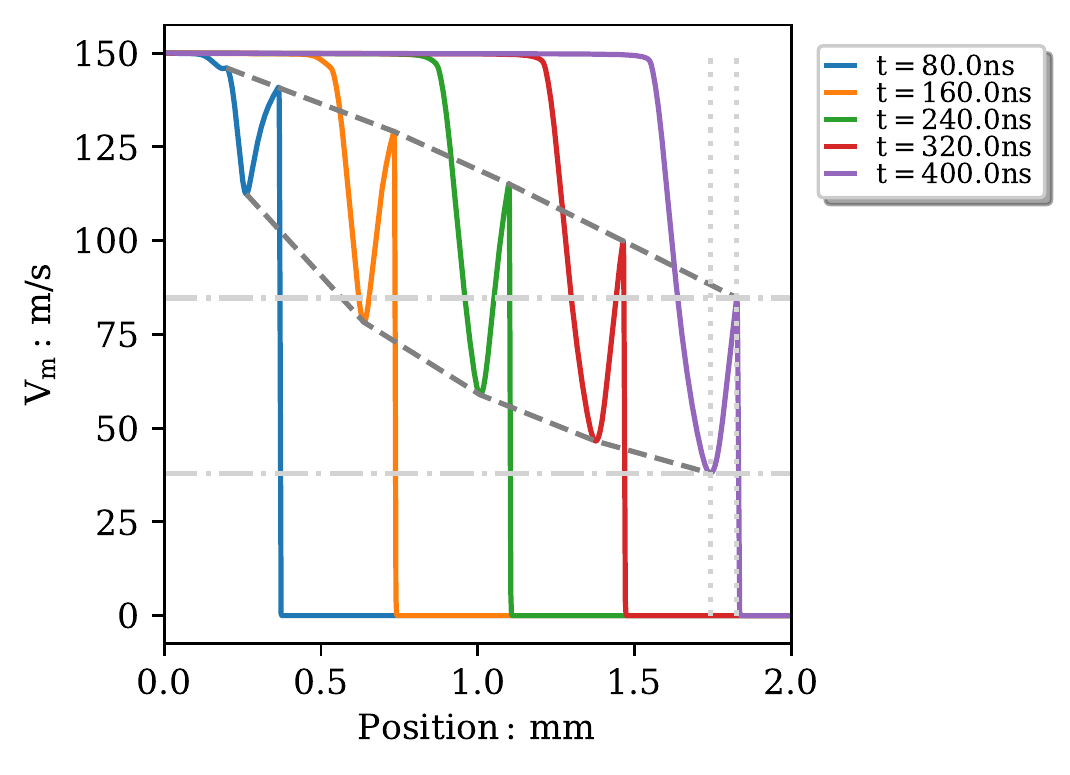}%
	\includegraphics[trim=0 0.3cm 0 0,clip,width=0.567\textwidth]{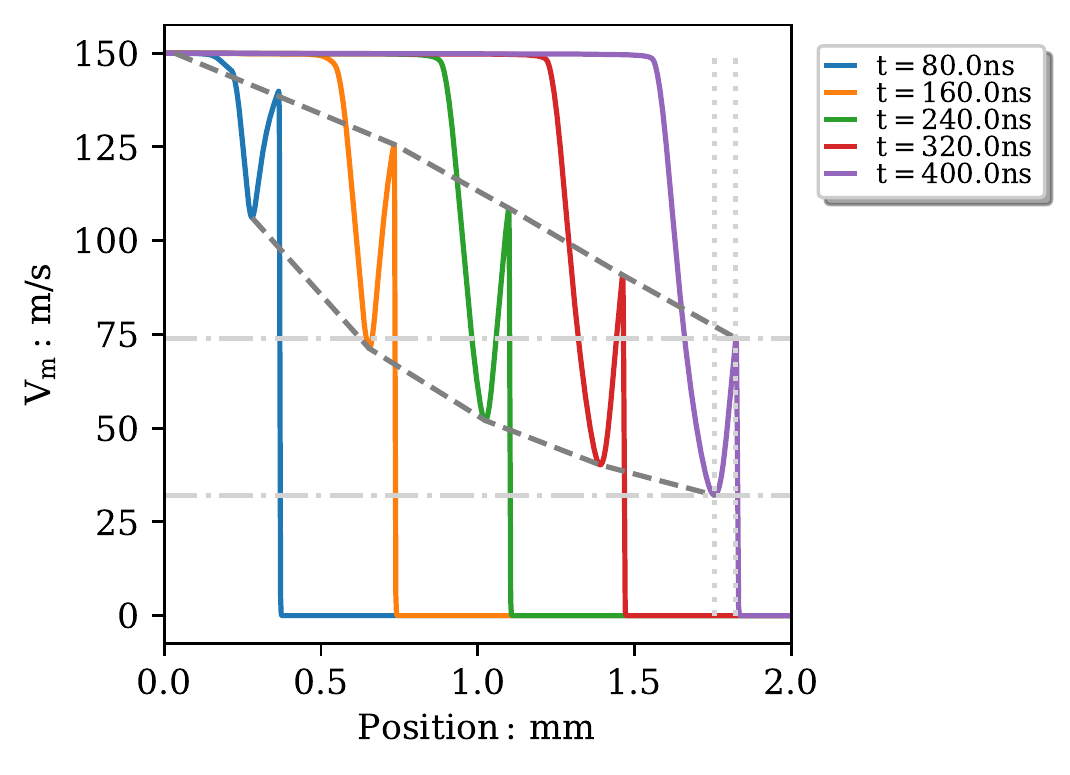}
	\caption{We show a sensitivity study of the elastic precursor decay in a 1D flyer plate impact simulation with one single crystal grain of copper taking into account different numbers of dislocation characters.
	On the top left we consider only edge dislocations,
	whereas on the top right, pure screw and edge dislocations are simulated.
	In the bottom row a total of 3 (left) and 7 (right) character angles, $\vth=(0,\pi/12,\pi/6,\pi/4,\pi/3,5\pi/12,\pi/2)$, are resolved.
		%The required input parameters $B_0(\vth) = (31, 33, 31, 30, 27, 25, 24)$ $\mu$Pas and $\cs(\vth)=(2.20, 1.77, 1.62, 1.77, 2.00, 1.77, 1.62)$ km/s were determined with PyDislocDyn \cite{pydislocdyn}.
		The total initial dislocation density $\varrho_\txt{disloc} =5\times10^6$ mm$^{-2}$ was the same in all four cases, but equally distributed among the resolved character angles (i.e. the number of edge dislocations was $1/7$ of this number for the bottom right etc.).
		The differences we see in precursor decay are due to differences in dislocation density evolution, see Figure \ref{fig:characterdensity}.
%		Unless noted otherwise, reference simulation parameters as listed in Table \ref{tab:inputdata} were used in all four cases.
	}
	\label{fig:character}
\end{figure}

Next, we study the importance of resolving many dislocation characters within a simulation.
Researchers often restrict the model to pure screw and edge (or even ignore this difference) in order to reduce computation time.
In anisotropic crystals, however, screw and edge dislocations do not decouple in the sense that neither the dislocation displacement field nor the drag coefficient of a mixed dislocation can be written as a linear superposition of pure edge and screw (as would be the case in the isotropic limit).
Therefore, it stands to reason that simulations could be improved by resolving also a number of mixed dislocations.
For simplicity, we assume here that the initial dislocation density is equally distributed among all resolved dislocation characters.
In Figure \ref{fig:character} we show how the elastic precursor decay changes when considering only pure edge dislocations (top left), screw and edge dislocations (top right), or three (bottom left) and seven (bottom right) dislocation characters.
Even though the total initial dislocation density is the same in all four simulations, the elastic precursor decays faster for seven than for one dislocation character.
%%% 
As we see below, reducing the initial dislocation density will counteract this effect.

\begin{figure}[!h!t]
	\centering
	\figtitle{Cu: dislocation character resolution affects dislocation density evolution}
	\includegraphics[trim=0 0.3cm 0cm 0,clip,width=0.45\textwidth]{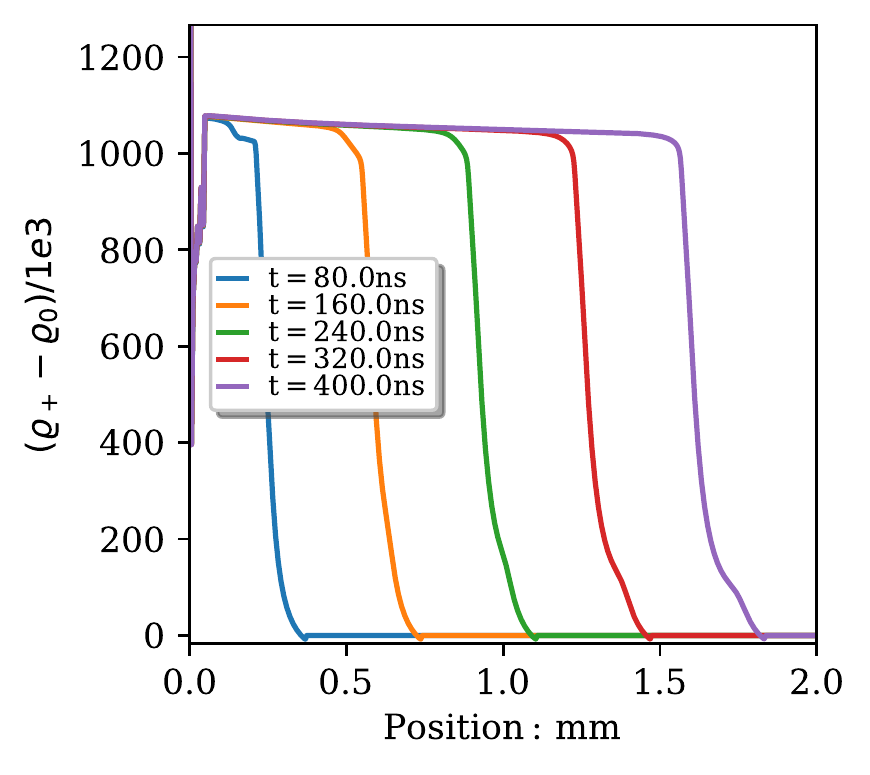}%
	\includegraphics[trim=0 0.3cm 0 0,clip,width=0.45\textwidth]{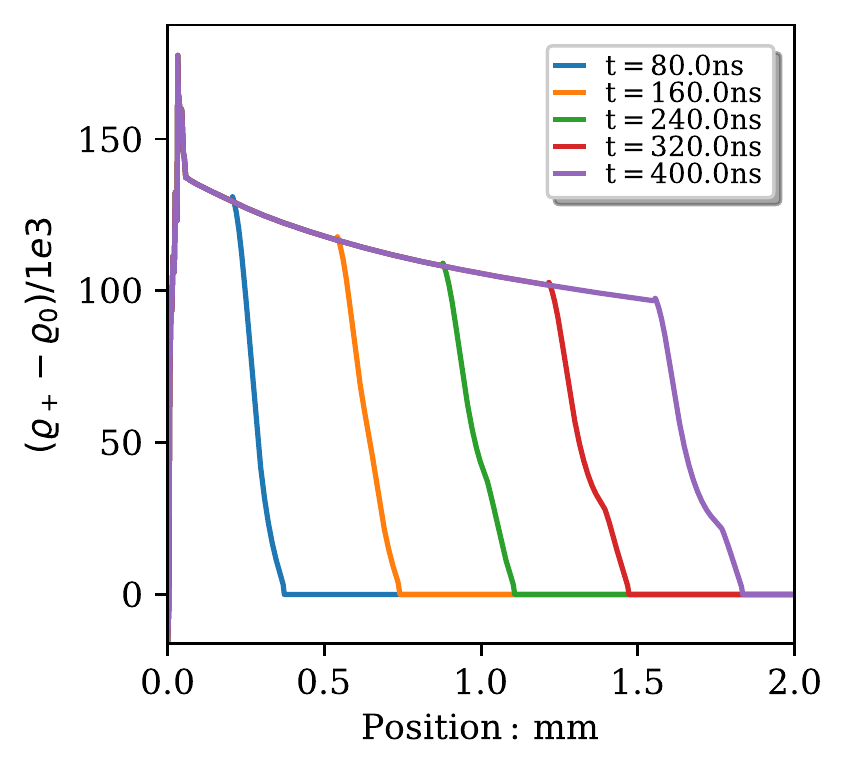}
	\includegraphics[trim=0 0.3cm 0cm 0,clip,width=0.45\textwidth]{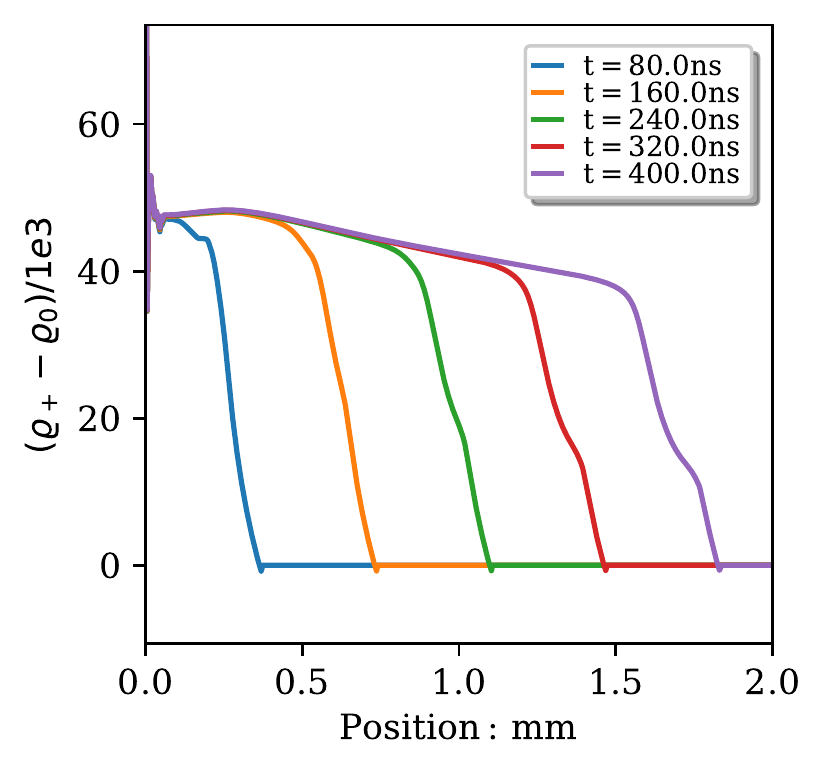}%
	\includegraphics[trim=0 0.3cm 0 0,clip,width=0.45\textwidth]{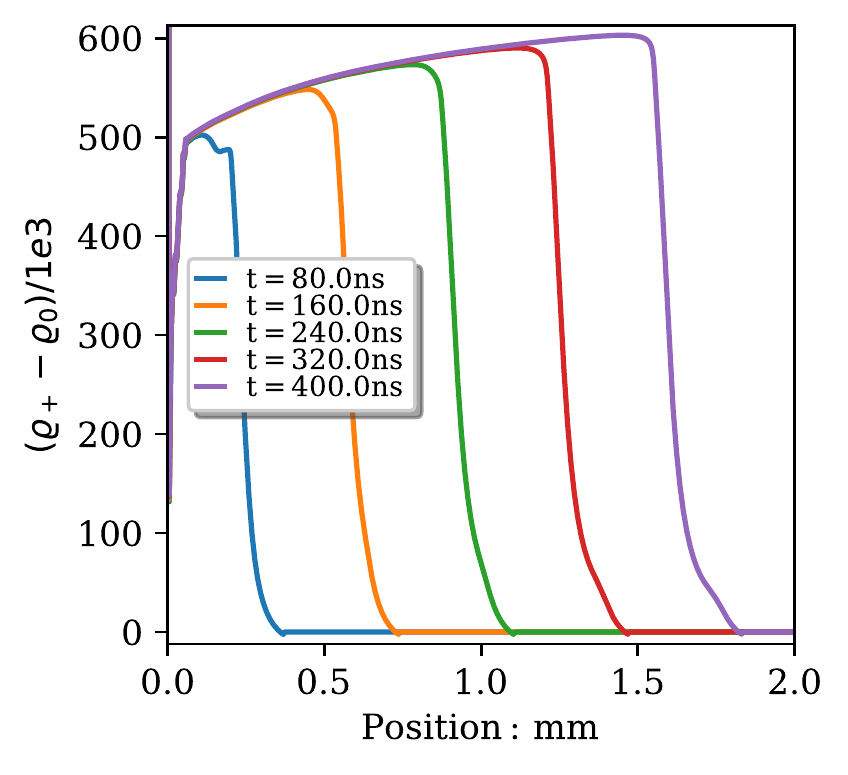}
	\caption{We show how the dislocation density evolution changes with the number of resolved dislocation characters.
		On the top left, only edge dislocations were simulated, whereas the other three plots show the density evolution of a simulation with 3 dislocation characters.
		In particular, the top right shows the density evolution for screw dislocations and the bottom row shows the same for character angle $\vth=\pi/4$ (left) and pure edge dislocations (right).
		These figures reveal that the total number of created dislocations during the simulation (due to multiplication and nucleation) is smaller in the 3-character case than in the pure edge case.
%		Unless noted otherwise, reference simulation parameters listed in Table \ref{tab:inputdata} were used in all four cases.
	}
	\label{fig:characterdensity}
\end{figure}

Drawing the reader's attention to Figure \ref{fig:characterdensity},  we see 
%the somewhat unexpected results 
that, interestingly, our dislocation source terms \eqref{eq:newsource_nonucl}, \eqref{eq:newsource_nucl} create fewer dislocations in total during the simulation if more character angles are considered.
In particular, this figure shows how the dislocation density evolution changes between resolving 1 character (top left) and 3 characters (where the top right shows $\vth=0$, bottom left shows $\vth=\pi/4$, and bottom right shows $\vth=\pi/2$).
%Interestingly, we see that our dislocation source terms \eqref{eq:newsource_nonucl}, \eqref{eq:newsource_nucl} create fewer dislocations in total during the simulation if more character angles are considered.
This can perhaps be understood by the following reasoning:
A more diverse (mixed) dislocation population is able to accommodate plastic flow more easily because then also their glide plane orientations are more diversely oriented with respect to the impact direction than if only edge dislocations are accounted for, and thus fewer additional dislocations need to be created in the process.

\begin{figure}[!h!t]
	\centering
	\figtitle{Cu: influence of impact orientation on precursor decay}
	\includegraphics[trim=0 0.3cm 2.585cm 0,clip,width=0.434\textwidth]{figures/copper/ref/CuRef.fig.dmb_vel_x1.pdf}%
	\includegraphics[trim=0 0.3cm 0 0,clip,width=0.567\textwidth]{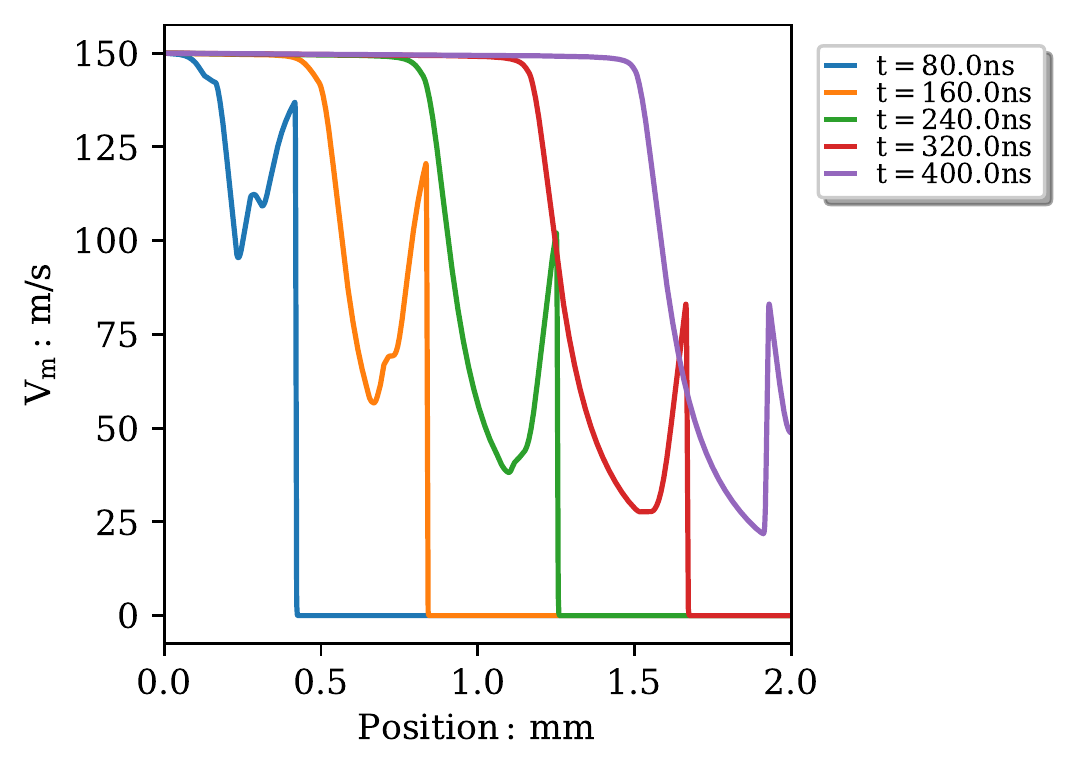}
	\includegraphics[trim=0 0.3cm 2.585cm 0,clip,width=0.434\textwidth]{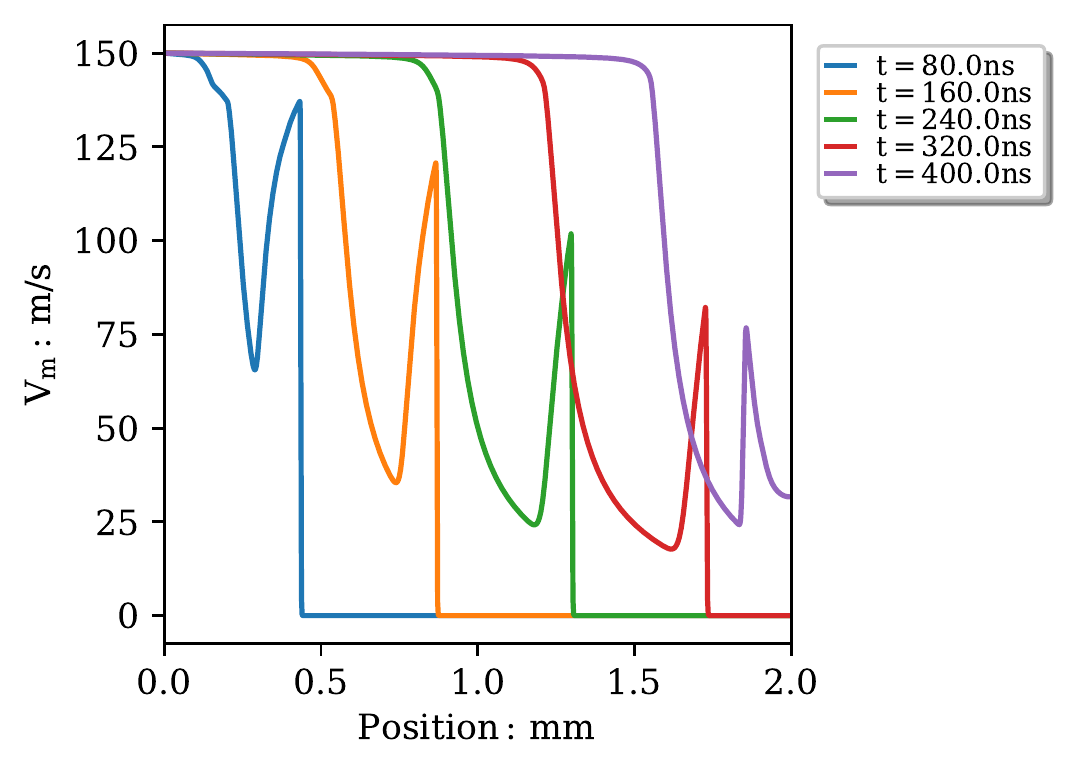}%
	\includegraphics[trim=0 0.3cm 0 0,clip,width=0.567\textwidth]{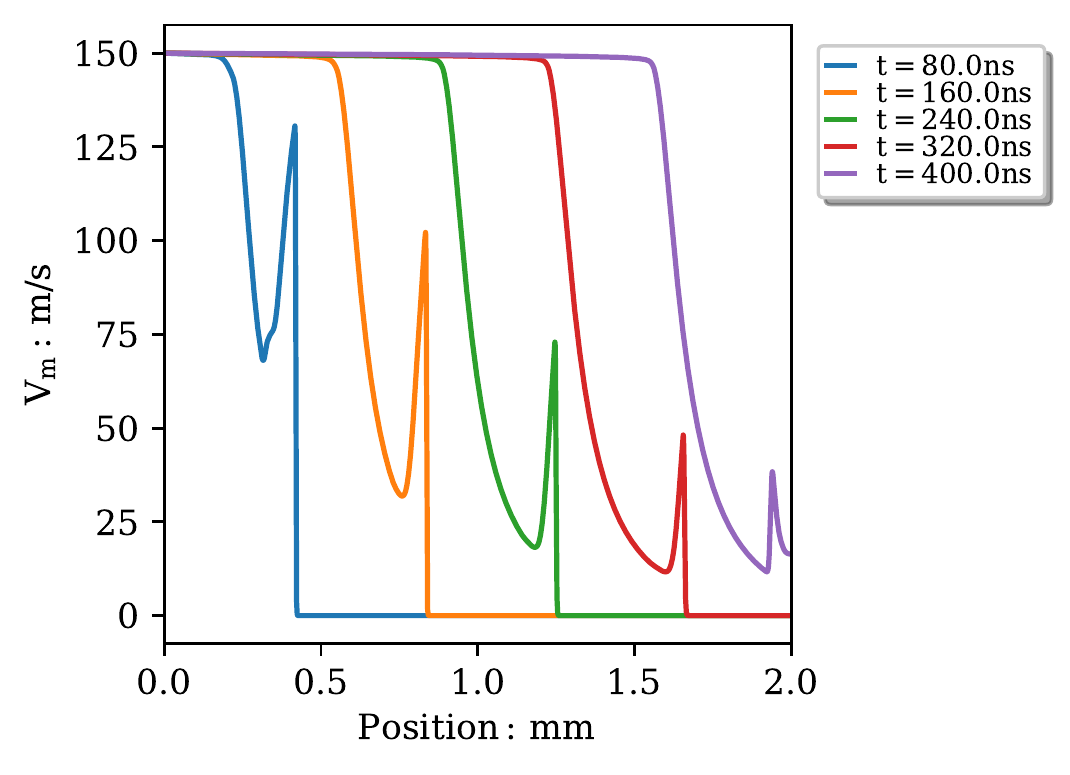}
	\caption{We show the dependence of the elastic precursor decay in a 1D flyer plate impact simulation on the orientation of the single crystal grain of copper.
		On the top left, the single crystal was impacted along one of the crystal axes.
		On the top right, the impact direction was aligned with one of the Burgers vectors.
		On the bottom left, the impact direction was aligned with one of the slip plane normals.
		The bottom right shows again the impact direction aligned with one of the Burgers vectors,
		but now with dislocation nucleation increased by an order of magnitude ($\dot{\varrho}_{n_0}=5\times10^{12}$ instead of $\dot{\varrho}_{n_0}=5\times10^{11}$).
%		Unless noted otherwise, reference simulation parameters listed in Table \ref{tab:inputdata} were used in all four cases.
	}
	\label{fig:orientation}
\end{figure}

The dependence on impact orientation, which we have seen in Figure \ref{fig:jones} above, is subsequently studied in more detail within Figure \ref{fig:orientation}:
%%%
The top left recaps once more the simulation with the impact direction aligned with a crystal axis.
The top right shows the same for an impact direction aligned with one of the Burgers vector directions whereas the bottom left shows the other extreme where the impact direction is aligned with one of the slip plane normals.
We see, as is expected, that the wave speeds are not the same in these three plots:
The longitudinal sound speed along the crystal axes is straightforwardly computed to be around 4.3 km/s in Cu with the elastic constants and material density of Table \ref{tab:inputdata}, whereas along a Burgers vector it is close to 5 km/s and along a slip plane normal it is even higher (hence the 400 ns wave in the lower left plot of Figure \ref{fig:orientation} has already been reflected from the 2 mm surface and is traveling to the left).

\begin{figure}[!h!t]
	\centering
	\figtitle{Cu: influence of impact orientation and drag coefficient on dislocation velocities}
	\includegraphics[trim=0 0.3cm 0 0,clip,width=0.45\textwidth]{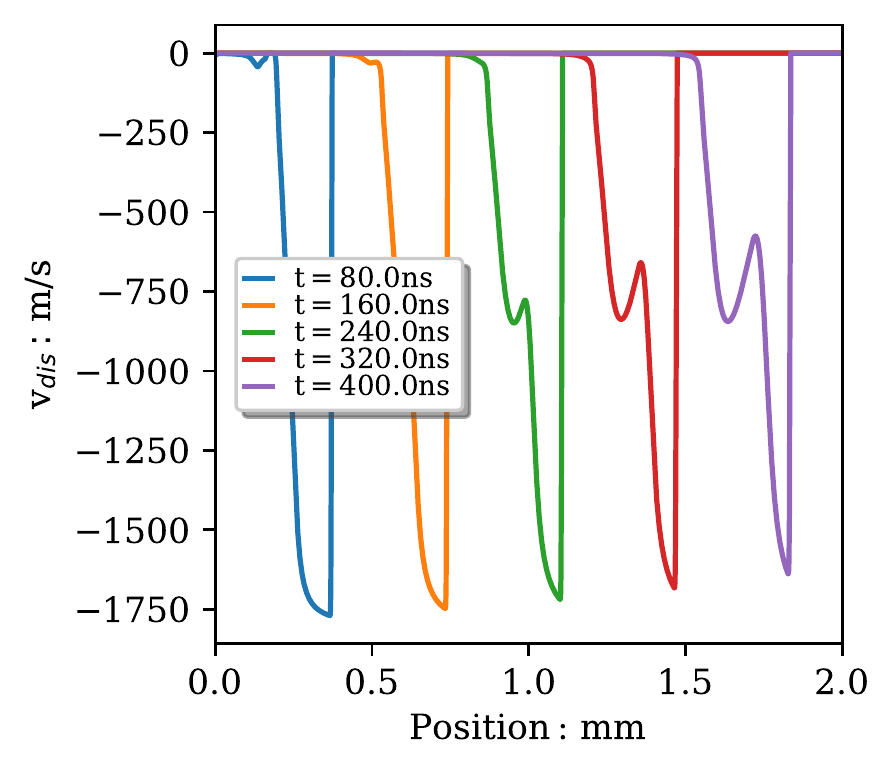}%
	\includegraphics[trim=0 0.3cm 0 0,clip,width=0.45\textwidth]{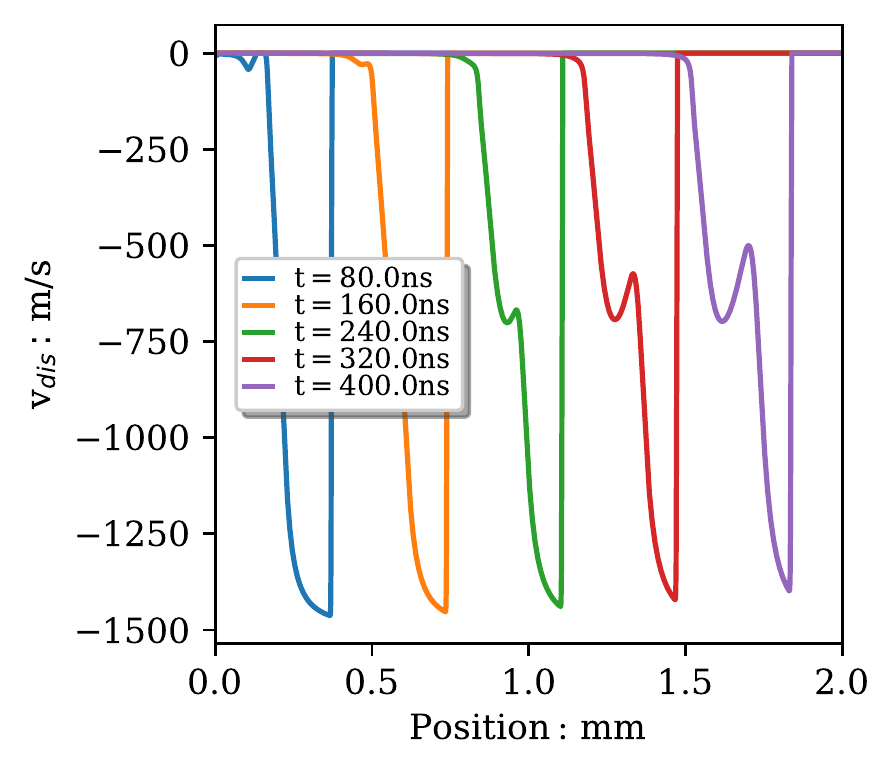}
	\includegraphics[trim=0 0.3cm 0 0,clip,width=0.45\textwidth]{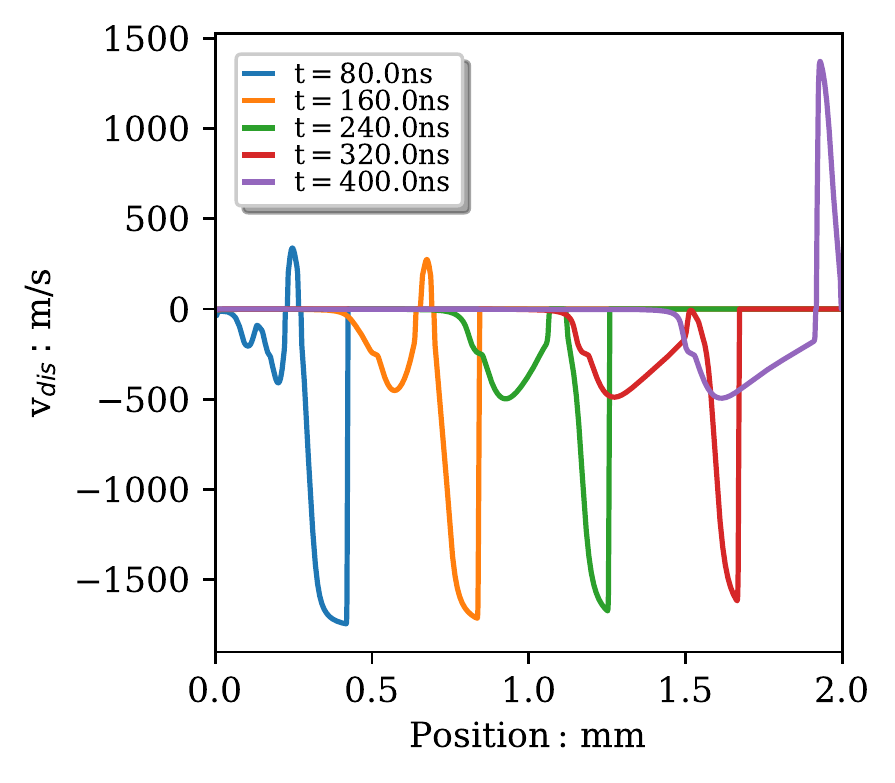}%
	\includegraphics[trim=0 0.3cm 0 0,clip,width=0.45\textwidth]{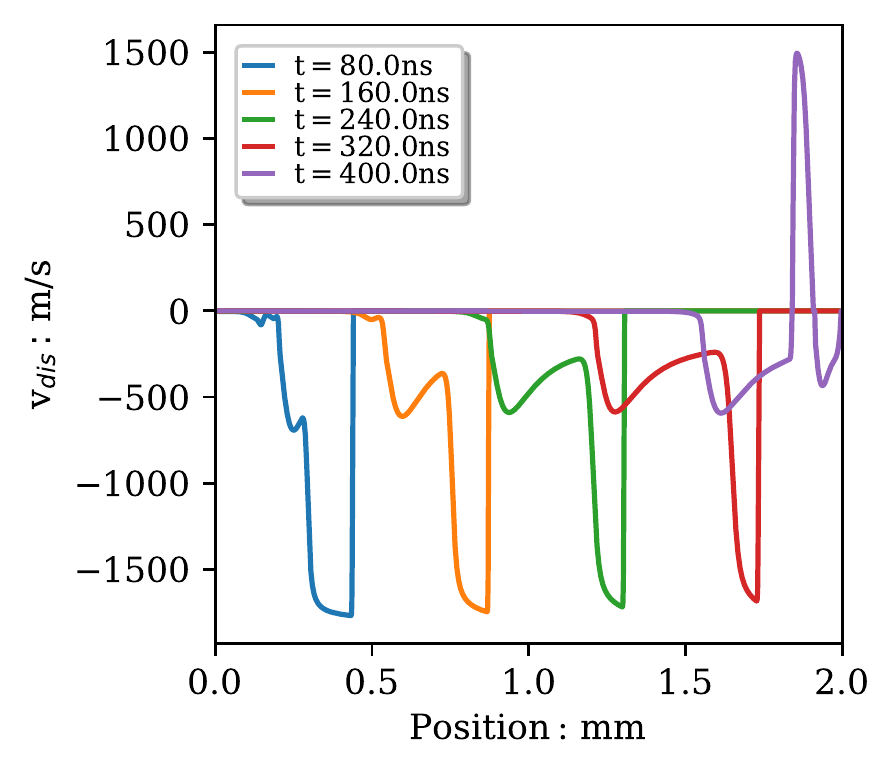}
	\caption{We show how the edge dislocation velocities are affected by the drag coefficient as well as the impact orientation.
		In particular, each of the four plots shows edge dislocation velocities for one of the active slip systems.
		In the top row we compare once more the two drag coefficients \eqref{eq:BTrho} (left) and \eqref{eq:BofsigmaAustin} (right) when the impact is aligned with one of the crystal axes.
		The bottom row shows once more the `reference' parameters (with drag \eqref{eq:BTrho}), but with the impact direction aligned with one of the Burgers vectors (left) and with one of the slip plane normals (right).
	}
	\label{fig:dis-vel-edge}
\end{figure}

Furthermore, depending on the alignment, not all slip planes are active, as can be easily worked out by projecting the stress computed via the stress-strain relations from the uniaxial strain caused by the flyer plate onto the respective slip systems.
In other words, dislocations in some of the 12 slip systems are not activated throughout the simulation due to the geometry of the impact.
At least a third to a half of the slip systems are however always active and cause the precursor decay.
Since on the top right, we see a small `double peak' at 80 nanoseconds which requires further studies, we repeat the same simulation with enhanced dislocation nucleation on the bottom right.
In the latter case we see no double peaks, i.e. they are washed out by the enhanced dislocation nucleation.
We therefore show the dislocation velocity magnitudes for one of the active slip systems in each case in Figure
%s \ref{fig:dis-vel-screw} (screw) and \ref{fig:dis-vel-edge} (edge).
\ref{fig:dis-vel-edge}  at the example of pure edge dislocations.
In particular, the top rows of both figures compare the difference between using the drag coefficient \eqref{eq:BTrho} from our first-principles derivation and approximate temperature / density dependence and using the ad-hoc empiric drag coefficient \eqref{eq:BofsigmaAustin}.
Both cases are once more aligned with a crystal axis.
In the bottom row, we considered \eqref{eq:BTrho}, but with the impact aligned with a Burgers vector (left) and a slip plane normal (right).
From the bottom left of this figure, we see that the double peak seen in the top right of Figure \ref{fig:orientation} is
% correlated with reverse dislocation motion.
associated with similar dislocation motion.
We emphasize that we get the same results independent of resolution in $x$ and/or time and also independent of the numerical scheme; for example repeating the calculation using the forward Euler or a fourth order Runge-Kutta scheme yields the same double-hump.
%Likewise, we find the same result using an upwind Petrov Galerkin method as with a finite volume update scheme using a local Lax-Friedrichs method.
See Ref. \cite{Mayeur:2016} for details on the implementations of various numerical schemes employed in our simulations.

\begin{figure}[!h!t]
	\centering
	\figtitle{Cu: sensitivity to initial dislocation density}
	\includegraphics[trim=0 0.3cm 2.585cm 0,clip,width=0.434\textwidth]{figures/copper/ref/CuRef.fig.dmb_vel_x1.pdf}%
	\includegraphics[trim=0 0.3cm 0 0,clip,width=0.567\textwidth]{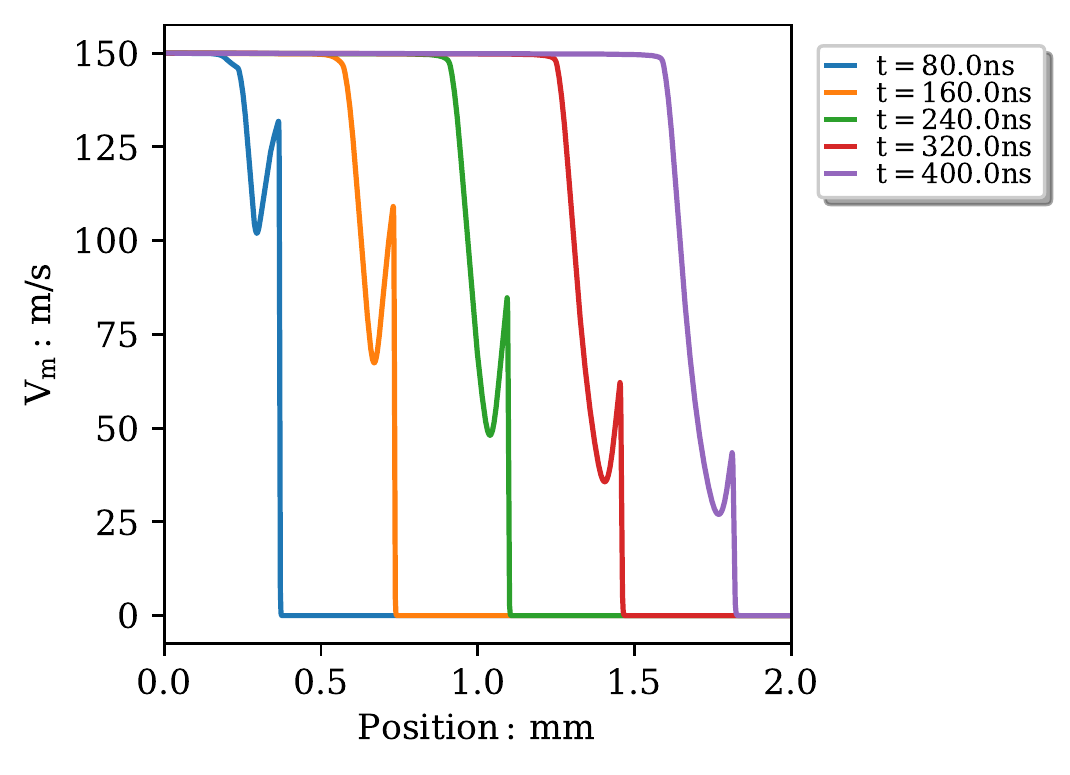}
	\includegraphics[trim=0 0.3cm 2.585cm 0,clip,width=0.434\textwidth]{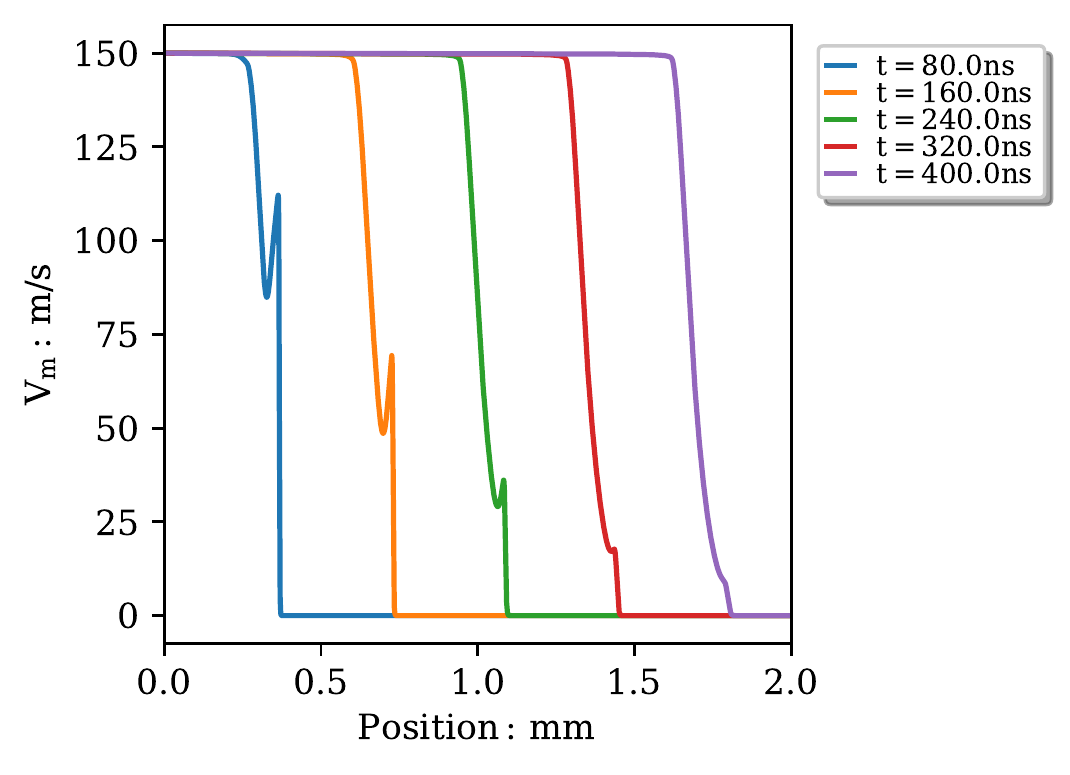}%
	\includegraphics[trim=0 0.3cm 0 0,clip,width=0.567\textwidth]{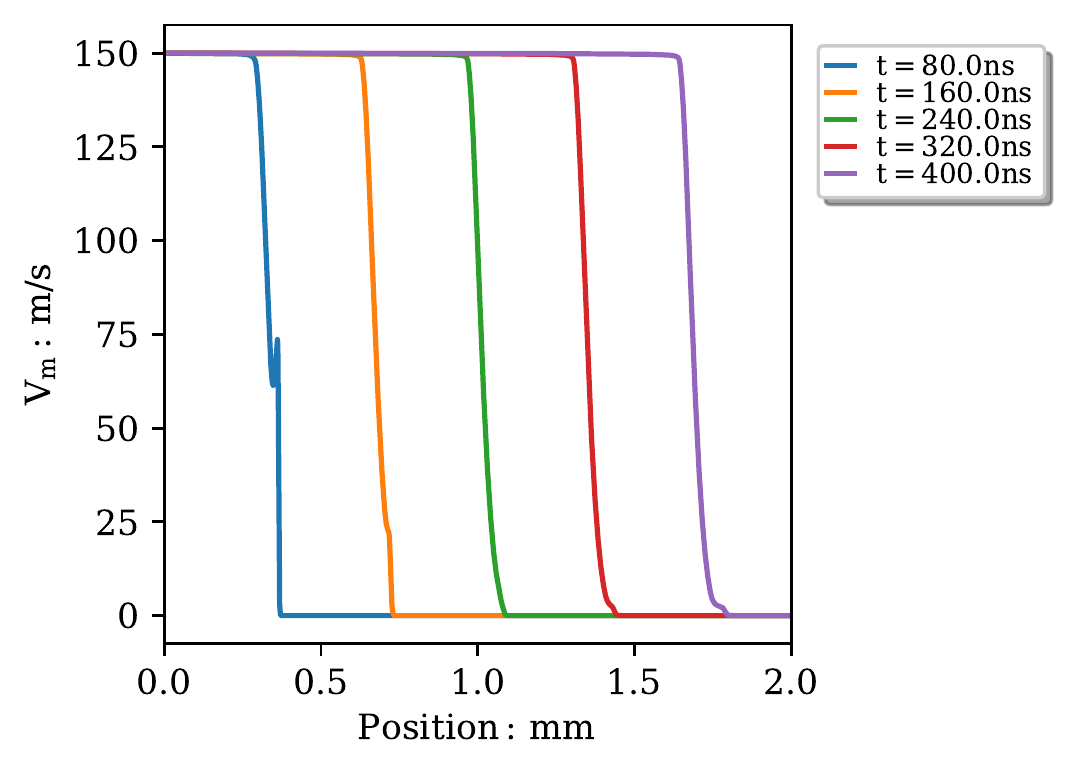}
	\caption{We show how sensitive the elastic precursor decay in a 1D flyer plate impact simulation with one single crystal grain of copper is with regard to the initial dislocation density.
		In the top row, we ran the simulation with $\varrho_\txt{disloc} =10^6$ [mm$^{-2}$] (left) and $\varrho_\txt{disloc} =2\times10^6$ [mm$^{-2}$] (right).
		In the bottom row we show the same for $\varrho_\txt{disloc} =4\times10^6$ [mm$^{-2}$] (left) and $\varrho_\txt{disloc} =8\times10^6$ [mm$^{-2}$] (right).
		Unless noted otherwise, reference simulation parameters as listed in Table \ref{tab:inputdata} were used in all four cases.}
	\label{fig:density}
\end{figure}

In Figure \ref{fig:density}, we show how sensitive the simulation is to the initial dislocation density:
Changing it by less than an order of magnitude can completely wash out the elastic precursor, i.e. the amplitudes are greatly diminished while their widths remain roughly the same.
This is particularly important, since at elevated temperatures one may expect the dislocation density to decrease due to enhanced annihilation prior to impacting the metal.

\begin{figure}[!h!t]
	\centering
	\figtitle{Cu: sensitivity to the dislocation multiplication rate}
	\includegraphics[trim=0 0.3cm 2.585cm 0,clip,width=0.434\textwidth]{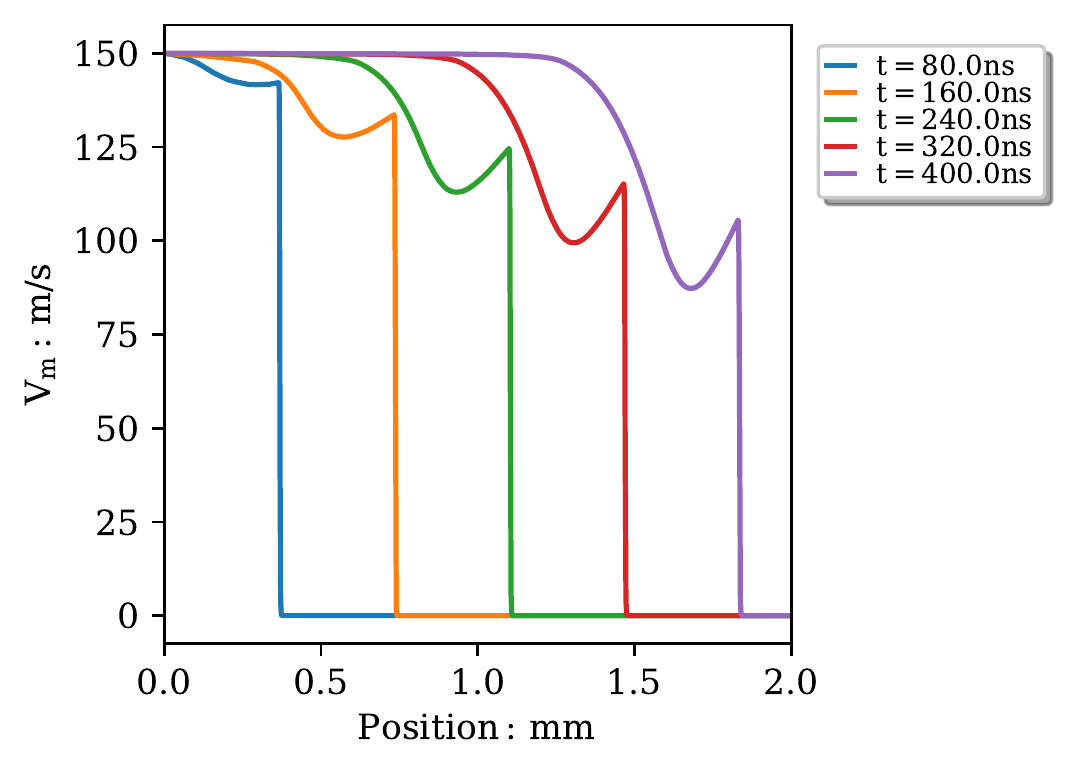}%
	\includegraphics[trim=0 0.3cm 0 0,clip,width=0.567\textwidth]{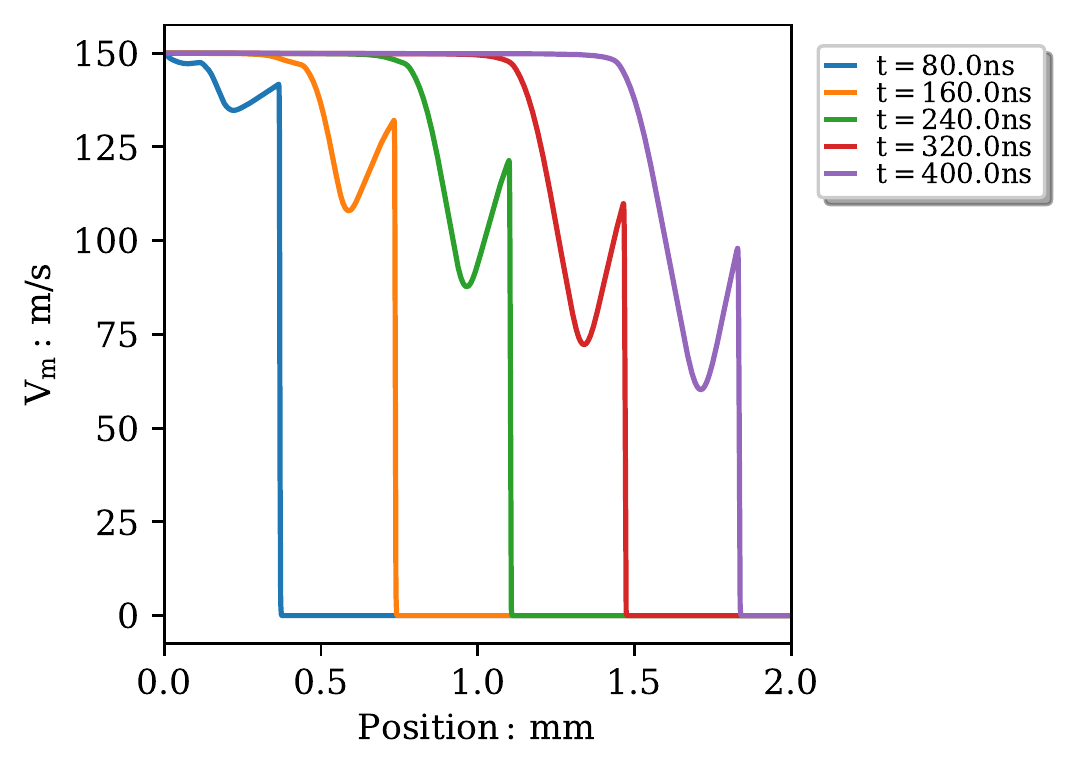}
	\includegraphics[trim=0 0.3cm 2.585cm 0,clip,width=0.434\textwidth]{figures/copper/ref/CuRef.fig.dmb_vel_x1.pdf}%
	\includegraphics[trim=0 0.3cm 0 0,clip,width=0.567\textwidth]{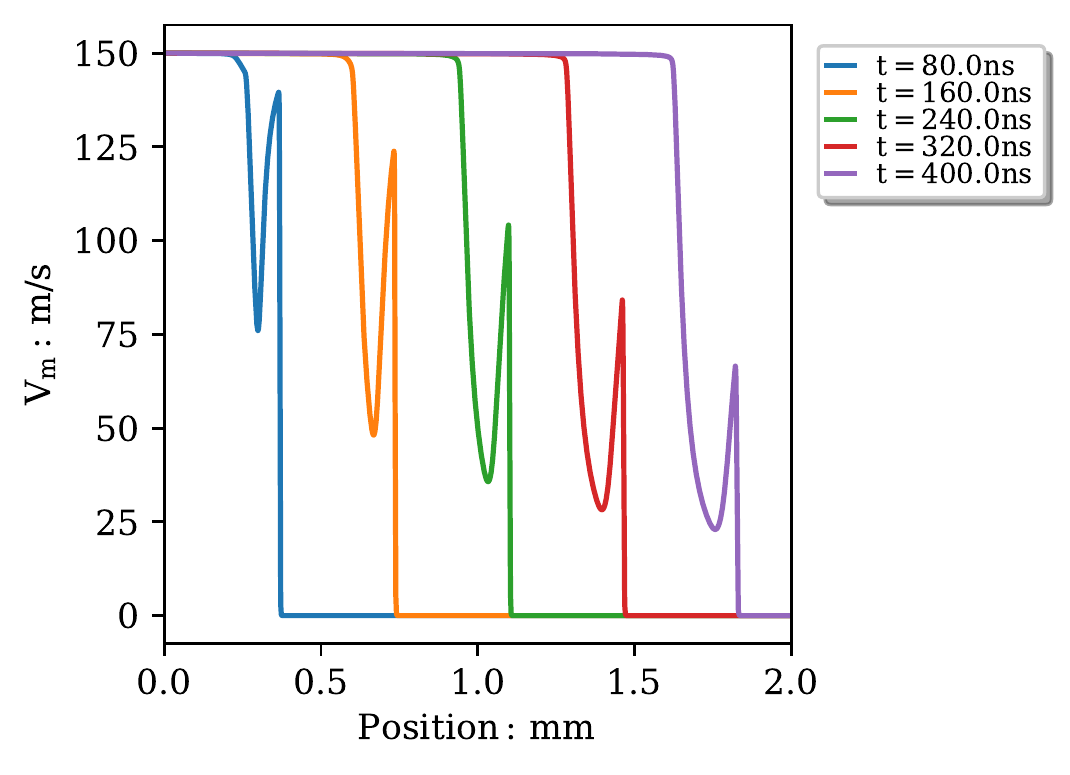}
	\caption{We show how sensitive the elastic precursor decay in a 1D flyer plate impact simulation with one single crystal grain of copper is with regard to dislocation multiplication.
		In the top row, we set the multiplication prefactor to $C_M=1/300$ (left) and $C_M=1/60$ (right).
		In the bottom row we show the same for $C_M=1/30$ (left) and $C_M=1/15$ (right).
		Unless noted otherwise, reference simulation parameters listed in Table \ref{tab:inputdata} were used in all four cases.}
	\label{fig:multiplication}
\end{figure}

Enhancing dislocation multiplication, as shown in Figure \ref{fig:multiplication}, yields faster elastic precursor decay and at the same time sharper (higher and narrower) spikes at early times.
Austin \cite{Austin:2018} also found enhanced dislocation multiplication to be correlated with faster elastic precursor decay and narrower spikes at early times, albeit with smaller amplitudes in his isotropic model in contrast to our present anisotropic model.
The faster decrease in amplitude of the elastic precursor with propagation distance is not surprising since an increased density of mobile dislocations relaxes the non-equilibrium stress more rapidly \cite{Austin:2018,Taylor:1965,Rohde:1969,Gillis:1971}.

\begin{figure}[!h!t]
	\centering
	\figtitle{Cu: sensitivity to the dislocation annihilation rate}
	\includegraphics[trim=0 0.3cm 2.585cm 0,clip,width=0.434\textwidth]{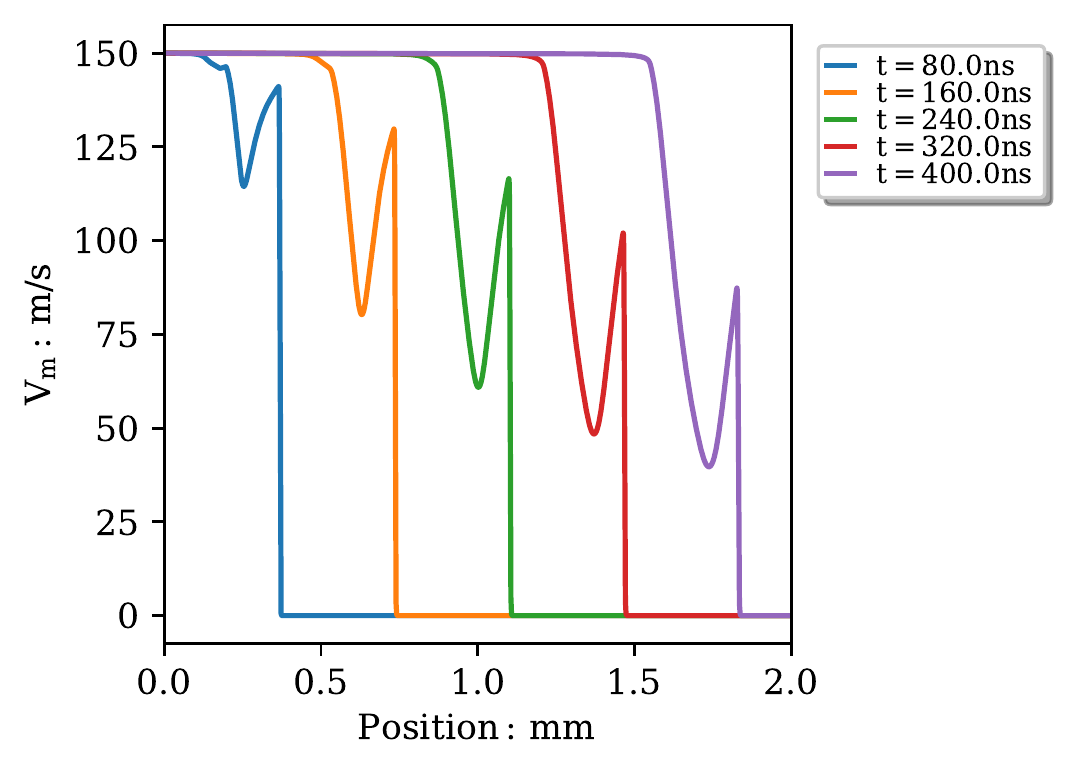}%
	\includegraphics[trim=0 0.3cm 0 0,clip,width=0.567\textwidth]{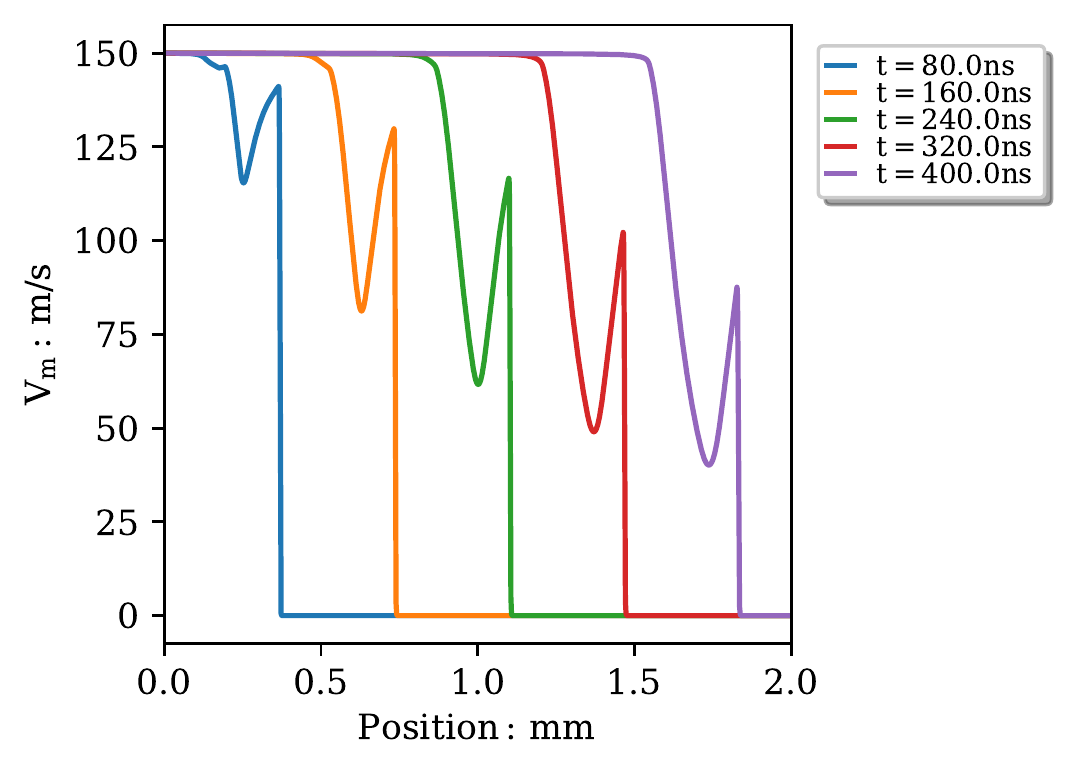}
	\caption{We show how sensitive the elastic precursor decay in a 1D flyer plate impact simulation with one single crystal grain of copper is with regard to dislocation annihilation.
		%In the top row, we set the annihilation prefactor to $C_A=0.5$ (left) and $C_A=2$ (right).
		%In the bottom row we show the same for $C_A=5$ (left) and $C_A=10$ (right).
		%Unless noted otherwise, reference simulation parameters are listed in Table \ref{tab:inputdata} were used in all four cases.
		On the right, we set the annihilation prefactor to $C_A=0.5$ and on the left to $C_A=10$.
		We note that the present simulation is not very sensitive to this parameter.
	}
	\label{fig:annihilation}
\end{figure}

Dislocation annihilation, on the other hand has hardly any effect during the impact simulation, as seen in Figure \ref{fig:annihilation},
consistent with the findings in Ref. \cite{Austin:2018}.

\begin{figure}[!h!t]
	\centering
	\figtitle{Cu: sensitivity to the dislocation nucleation rate}
	\includegraphics[trim=0 0.3cm 2.585cm 0,clip,width=0.434\textwidth]{figures/copper/ref/CuRef.fig.dmb_vel_x1.pdf}%
	\includegraphics[trim=0 0.3cm 0 0,clip,width=0.567\textwidth]{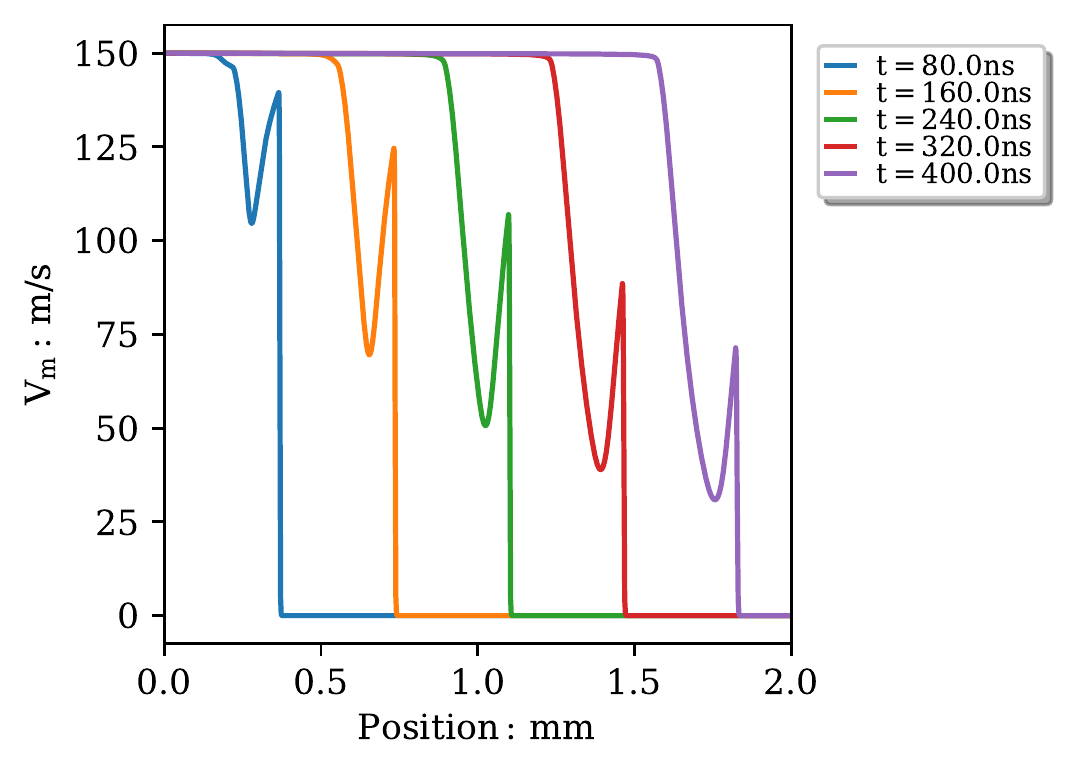}
	\includegraphics[trim=0 0.3cm 2.585cm 0,clip,width=0.434\textwidth]{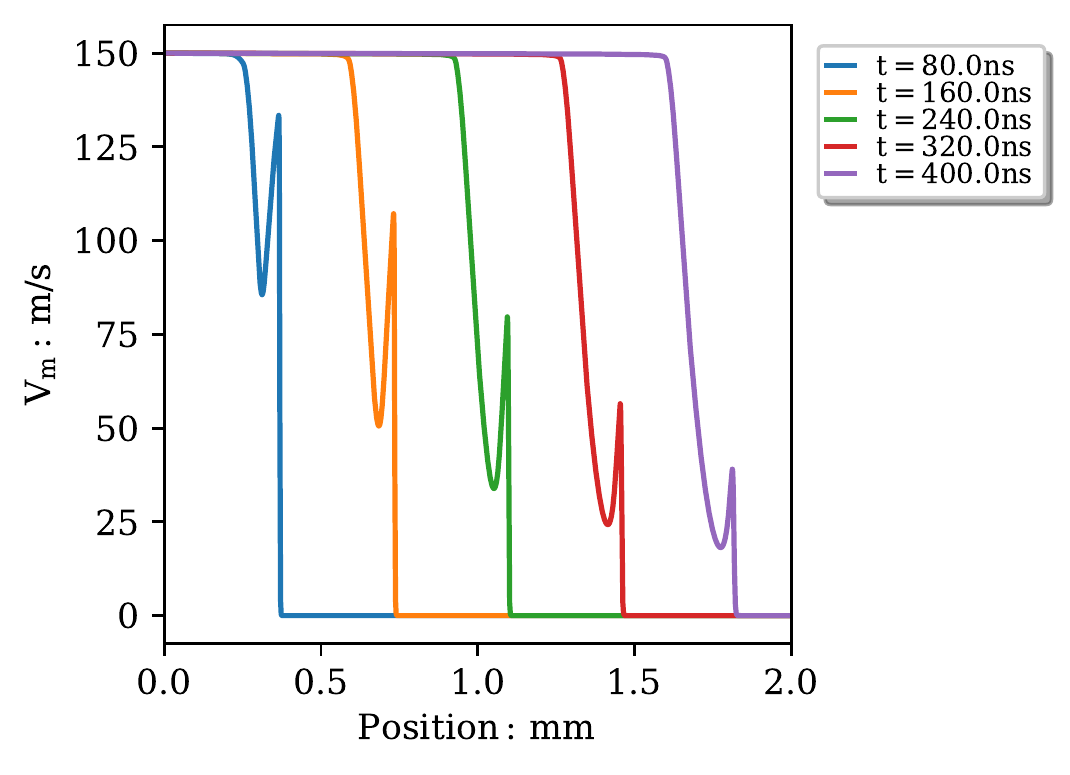}%
	\includegraphics[trim=0 0.3cm 0 0,clip,width=0.567\textwidth]{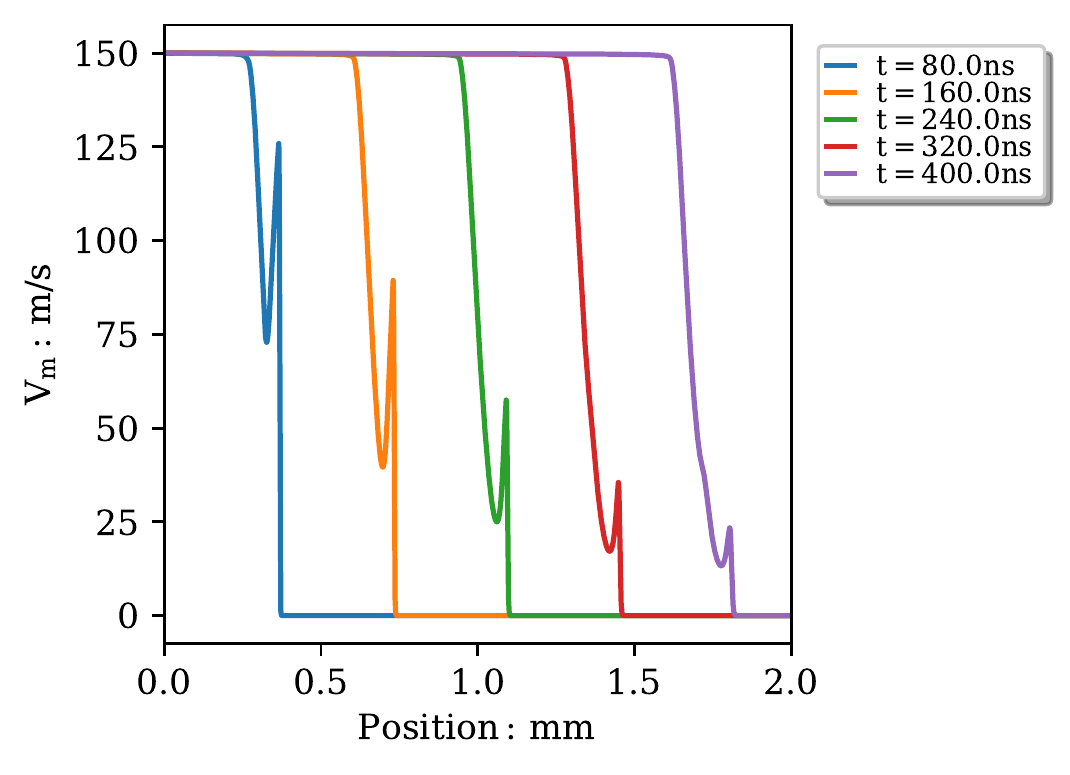}
	\caption{We show how sensitive the elastic precursor decay in a 1D flyer plate impact simulation with one single crystal grain of copper is with regard to the rate of dislocation nucleation.
		In the top row, we set the maximum nucleation rate to $\dot{\varrho}_{n_0}=5\times10^{11}$ (left) and $\dot{\varrho}_{n_0}=2\times10^{12}$ (right).
		In the bottom row we show the same for $\dot{\varrho}_{n_0}=8\times10^{12}$ (left) and $\dot{\varrho}_{n_0}=1.6\times10^{13}$ (right).
		Unless noted otherwise, reference simulation parameters listed in Table \ref{tab:inputdata} were used in all four cases.}
	\label{fig:nucleation}
\end{figure}

The simulation is, however, very sensitive to dislocation nucleation, as seen in Figure \ref{fig:nucleation}:
Enhanced nucleation reduces the width and height of the spikes and leads to faster precursor decay.
The same conclusion was drawn in Ref. \cite{Austin:2018} within an isotropic model.

\section{Discussion}
\label{sec:comparison}
%%%%%%%%%%%%%%%%%%%%%%%%%%%%%%%%%%

In Ref. \cite{Austin:2018} it was speculated that the increase in elastic precursor spike amplitude at high temperatures observed in poly-crystalline aluminum experiments \cite{Zaretsky:2012} is due to the increased drag from phonon scattering.
Similar speculations were put forward to explain single crystal aluminum experiments in Ref. \cite{Kanel:2001}, but the later Ref. \cite{Zaretsky:2012} indeed pointed out the important role of dislocation density evolution (in addition to dislocation drag) for the temperature dependence of elastic precursor decay.
In short, there has been a debate in the literature as to which of the two effects is more important.

The results of our present simulations, using a temperature, pressure, and character dependent dislocation drag coefficient derived from first principles, 
%are consistent with the premise that dislocation drag alone is not enough to explain the temperature dependence of elastic precursor decay in single crystal aluminum.
show that the temperature dependence of dislocation drag does indeed lead to a small increase of the elastic precursor amplitude in single crystal aluminum, but the effect is much smaller than what has been observed in experiments \cite{Zaretsky:2012,Austin:2018,Zuanetti:2021}.
Since we used the most accurate dislocation drag model to date, its influence on this effect could be quantified fairly well.
We therefore argue that the temperature dependence of dislocation density evolution, which is currently based on numerous model parameters whose temperature dependence is not well established, plays an even more important role.
%
%Of course our present simulations are for single crystal aluminum (not a poly-crystal), and for the temperature dependence of the elastic constants we used a very rough approximation.
%Nonetheless, the general trend of the drag coefficient is still the same, and there is no reason why in a polycrystal dislocation drag should suddenly lead to enhanced precursor spike amplitudes.

%In particular, while the dislocation annihilation parameter has a negligible effect, enhancing the dislocation multiplication by a moderate amount will lead to narrower spikes and faster precursor decay.
%While we currently lack a good dislocation density evolution model accounting for all these competing effects, we do see that the experimentally observed temperature dependence of the elastic precursor decay reported in Refs. \cite{Zaretsky:2012,Austin:2018,Zuanetti:2021} must have to do with a combination of dislocation density evolution and the drag coefficient.
Most importantly, we deduce that most likely the dislocation nucleation (and multiplication) mechanism is markedly different at high temperatures in aluminum, assuming that the initial dislocation density of an annealed specimen is roughly the same for different temperatures.
This view is further supported by recent results in polycrystalline aluminum of Zuanetti et al. \cite{Zuanetti:2021} who show that the inclusion of a temperature dependent dynamic recovery function controlling the rate of immobilization of dislocations at high plastic strains (which allows saturation of the dislocation density) improves the agreement between experimental and simulation results of elastic precursor decay at elevated temperatures.
A generalization of this recovery function to single crystal simulations is left for future work.

\subsection*{Acknowledgments}
%%%%%%%%%%%%%%%%%%%%%%%%%%%%%%%%%%%%%%%%%%
\noindent
D.J. L. is grateful for many relevant discussions with Ryan Austin, Nathan Barton, and Bryan Zuanetti.
We also thank the anonymous referees for their insightful comments.
This work was performed under the auspices of the U.S. Department of Energy under contract 89233218CNA000001.
In particular, the authors are grateful for the support of the Physics and Engineering Models sub-program element of the Advanced Simulation and Computing program.

%%%%%%%%% SENSITIVITY STUDY AT EXAMPLE OF COPPER: %%%%%%%%%%%%%

%\begin{figure}[!h!t]
%	\centering
%	\includegraphics[width=0.5\textwidth]{figures/copper/ref/CuRef.fig.dis_vel_s1.1.pdf}%
%	\includegraphics[width=0.5\textwidth]{figures/copper/drag/CuBaustin.fig.dis_vel_s1.1.pdf}
%	\includegraphics[width=0.5\textwidth]{figures/copper/orientation/CuRotBurg_s1.fig.dis_vel_s2.1.pdf}%
%	\includegraphics[width=0.5\textwidth]{figures/copper/orientation/CuRotNorm_s1.fig.dis_vel_s4.1.pdf}
%	\caption{We show how the screw dislocation velocities are affected by the drag coefficient as well as the impact orientation.
%		In particular, each of the four plots shows screw dislocation velocities for one of the active slip systems.
%		In the top row we compare once more the two drag coefficients \eqref{eq:BTrho} (left) and \eqref{eq:BofsigmaAustin} (right) when the impact is aligned with one of the crystal axes.
%		The bottom row shows once more the `reference' parameters (with drag \eqref{eq:BTrho}), but with the impact direction aligned with one of the Burgers vectors (left) and with one of the slip plane normals (right).
%	}
%	\label{fig:dis-vel-screw}
%\end{figure}

%%%%%%%%%%%%%%%%%%%%%%%%%%%%%%%%%%%%%%%%%%%
\bibliographystyle{utphys-custom}
\bibliography{dislocations}

\end{document}